\documentclass[aps,pre,a4paper,amsmath,amssymb,floatfix,twocolumn]{revtex4-1}
\usepackage{amsmath}         % features for displayed equations
\usepackage{amssymb}         % additional maths symbols
\usepackage{amsfonts}        % additional maths fonts
\usepackage{amsthm}          % theorem environments setup
\usepackage{xcolor}
\usepackage[final]{hyperref} % hypertext links
\usepackage{graphicx}        % extended graphics package
\usepackage{tabularx}        % better tables
\usepackage{booktabs}        % for better rules in tables
\usepackage{textcase}        % for \MakeTextUppercase
\usepackage[UKenglish]{babel}% correct hyphenation rules for UK English
\usepackage[T1]{fontenc}     % better treatment of accented words
\usepackage{multirow,bigdelim}
\usepackage[nomargin,inline,draft]{fixme}
\graphicspath{{figures/}}    % store figures in /figures/ subdirectory
\newcommand{\Tr}{\text{Tr}}							%Trace%
\,							%Infinitesimal, straight 'd'%
\newcommand{\set}[1]{ \left\lbrace #1 \right\rbrace }	%Set%
\newcommand{\bra}[1]{\langle #1 \vert}
\newcommand{\ket}[1]{\vert #1 \rangle}

\newcommand{\mat}[4]{\left(\begin{array}{cc} #1 & #2 \\ #3 & #4 \end{array}\right)}%
\theoremstyle{definition}

\fxsetup{theme=color,mode=multiuser}
\FXRegisterAuthor{mh}{amh}{\color{red}MvH}

\begin{document}
%
%
%
%%% PREAMBLE %%%
%
\title{Open quantum reaction-diffusion dynamics: absorbing states and relaxation}
\author{Merlijn van Horssen}
\author{Juan P. Garrahan}
\affiliation{School of Physics and Astronomy, University of
Nottingham, Nottingham, NG7 2RD, UK}

\begin{abstract}
We consider an extension of classical stochastic reaction-diffusion (RD) dynamics to open quantum systems.  We study a class of models of hard core particles on a one-dimensional lattice whose dynamics is generated by a quantum master operator and where particle hopping is coherent while reactions, such as pair annihilation or pair coalescence, are dissipative.  These are quantum open generalisations of the $A+A \to \varnothing$ and $A+A \to A$ classical RD models.  We characterise the relaxation of the state towards the stationary regime via a decomposition of the system Hilbert space into transient and recurrent subspaces.  
We provide a complete classification of the structure of the recurrent subspace (and the non-equilibrium steady states) in terms of the dark states associated to the quantum master operator and its general spectral properties.  We also show that, in one dimension, relaxation towards these absorbing dark states is slower than that predicted by a mean-field analysis due to fluctuation effects, in analogy with what occurs in classical RD systems.  Numerical simulations of small systems suggest that the decay of the density in one dimension, in both the open quantum $A+A \to \varnothing$ and $A+A \to A$ cases, may go asymptotically as $t^{-b}$ with $1/2 < b < 1$.  \end{abstract}
\date{\today}
\maketitle
%
%
%
%%% BODY %%%
%

\section{Introduction}

Non-equilibrium statistical mechanics is a field characterised by a richness of dynamical phenomena, many of which elude full theoretical understanding. One of the foremost theoretical challenges is the characterisation of non-equilibrium steady states (NESS) (and the relaxation towards them), which remains a topic of current research efforts \cite{Hinrichsen2000,Zia2007,Peliti2011}. Several classes of systems (such as particle hopping models and directed percolation) exhibit phase transitions from fluctuating phases into a particular type of NESS, namely absorbing states; once reached, such absorbing states cannot be left.  This behaviour is typical of  reaction-diffusion (RD) models \cite{Toussaint1983,Hinrichsen2000,Tauber2005}. If the diffusive mixing is not strong enough, asymptotic decay of global degrees of freedom can be slower than predicted by mean-field approximations \cite{Toussaint1983,Hinrichsen2000,Tauber2005}, behaviour which is explained by fluctuation effects which have been confirmed experimentally \cite{Kroon1993,Allam2013}, and continues to be a topic of current research \cite{Durang2014}.

In this paper, we will consider a type of dynamics similar to RD models, extended to an open quantum spin chain. The theory of open quantum systems \cite{Buchleitner2002, Gardiner2004} is the topic of ongoing current theoretical and experimental research in both quantum optics and cold atomic systems \cite{Diehl2008, Diehl2010a, Torre2012, Lee2012, Foss-Feig12, Ates2012a, Pichler10}. We consider a class of one-dimensional open quantum systems with dynamics analogous to that of classical RD models: particle propagation is coherent, through quantum hopping, while reactions between particles are dissipative.  We show that, as in the classical RD models, these quantum systems exhibit non-exponential decay to absorbing stationary states, with mean-field approximations to the dynamics failing to predict the correct rate of decay. We connect this behaviour to the algebraic structure of the system Hilbert space \cite{Baumgartner2008a} and present a classification of the dark states which generate the absorbing part of the dynamics -- thus fully describing the stationary states (or NESS) for the associated quantum master operator.  We also numerically study the dynamics of small systems via quantum jump Monte Carlo simulations \cite{Plenio1998} and find evidence for a power law decay of the particle density towards the absorbing state, with an exponent which appears to be neither that predicted by mean-field analysis, nor that of the classical one-dimensional RD systems. 

This article is organised as follows. We start by introducing the quantum reaction-diffusion models studied in this paper and discuss their immediate properties (explaining why they are sensible open quantum analogues of classical RD systems), by considering conservation of particle number and invariant subspaces, and by looking at their quantum jump trajectories. This first section concludes with a decomposition of the Hilbert space of these systems into transient (decaying) and recurrent (absorbing) subspaces.
The next section contains our main analytical result as we consider the recurrent subspace in more detail.  We provide a full classification of the dark states associated to our quantum RD models, and we consider the spectral properties of the quantum master operator to argue that these dark states generate the recurrent subspace.  We will also explain the asymptotic behaviour of the evolution of the quantum state in terms of these dark states; we discuss how not all of the recurrent subspace is within reach from any initial state.  We also provide an argument for the equivalence between annihilation and coagulation dynamics in our class of quantum RD models. The nontrivial asymptotic structure of the state resulting from the non-trivial collection of dark states is a fundamental feature of the quantum model, compared to classical RD dynamics.
In the final section we characterise the decay of the density of particles.  We show that a mean-field approximation to the dynamics, as in the classical RD case, predicts asymptotically a $t^{-1}$ decay which is reaction-limited.  In contrast, from quantum jump Monte Carlo simulations on small systems, we find evidence that in dimension one the dynamics is instead hopping-limited (cf.\ diffusion-limited) 
with a power law decay with an exponent smaller than the mean-field one, but apparently larger than that of the classical RD dynamics.

\section{Models}

In this section we introduce the family of physical models considered in this paper, along with prerequisite mathematical background. We discuss features of the dynamics of the models by considering associated quantum jump trajectories, and we characterise the structure of the system Hilbert space according to conservation of the number of particles.

We consider quantum models which are analogous to classical one-dimensional RD models.  Each of the models consist of a one-dimensional lattice where a site can be either empty, denoted by $0$, or occupied by a single particle, denoted by $1$.  Particles can hop between lattice sites, symbolically $(1,0) \leftrightarrow (0,1)$. A reaction may only occur when two adjacent sites are occupied. We will consider three particular types of reaction: analogous to the classical $A+A \to \varnothing$ reaction, we define \emph{pair annihilation}, where two neighbouring particles annihilate, symbolically denoted by $(1,1) \rightarrow (0,0)$. We also define two analogous reactions to the classical $A+A \to A$ reaction: \emph{asymmetric coagulation}, where two neighbouring particles coalesce into one, denoted by $(1,1) \rightarrow (1,0)$; and \emph{symmetric coagulation}, where two neighbouring particles coalesce in two possible ways, $(1,1) \rightarrow (1,0), (0,1)$.  The quantum nature of our models is given by the fact that we will take particle hopping to be coherent, while reactions will be dissipative.

With periodic boundary conditions the Hilbert space $\mathcal{H}$ of these models is that of a quantum spin chain of $N$ sites, $\mathcal{H} = \mathbb{C}^{2} \otimes \ldots \otimes \mathbb{C}^{2}$. The coherent part of the dynamics is quantum hopping with rate $\Omega>0$, described by a Hamiltonian
\begin{equation}
    H = \Omega \sum_{i=1}^{N} \left( \sigma_{i}^{-} \sigma_{i+1}^{+} + \sigma_{i}^{+} \sigma_{i+1}^{-} \right)
    \label{H}
\end{equation}
where $\sigma_{i}^\pm$ are Pauli operators acting on site $i$, and $\sigma_{N+1}^{\pm} \equiv \sigma_{1}^{\pm}$ due to the choice of periodic boundaries.  For any one-site operator $X$, the notation $X_{i}$ is employed as the usual shorthand for the tensor product $\mathbf{1}^{\otimes (i-1)} \otimes X \otimes \mathbf{1}^{\otimes (N-i)}$, where $\mathbf{1}$ denotes $2 \times 2$ identity matrix. We work in a basis for $\mathbb{C}^{2}$ in which the raising and lowering operators $\sigma^{+}$ and $\sigma^{-}$ take the form
\begin{equation*}
    \sigma^{+} = \mat{0}{1}{0}{0},\quad \sigma^{-} = \mat{0}{0}{1}{0}
\end{equation*}
with eigenvectors denoted by $\sigma^{-} \ket{0} = 0, \sigma^{-} \ket{1} = \ket{0}$; we denote the site number operator by $n = \sigma^{+} \sigma^{-}$.
    
The reactions are dissipative and the resulting dynamics of our model are those of an open quantum system.  In the Markov approximation, the corresponding time evolution of the density matrix is determined by a quantum master equation (QME), 
\begin{eqnarray}
\label{eq:masterequation}
\partial_t \rho &=& 
    \mathbb{W}(\rho) 
\\
    &=& -i \left[ H, \rho \right] + \sum_{\mu} \left( L_{\mu} \rho L_{\mu}^{\dagger} - \tfrac{1}{2} \left\lbrace L_{\mu}^{\dagger} L_{\mu}, \rho \right\rbrace \right).
    \nonumber
\end{eqnarray}
We use the notation $\left\lbrace \mathcal{T}_{t} := \exp t \mathbb{W}\right\rbrace_{t \geq 0}$ for the quantum dynamical semigroup giving the solution to the QME as $\rho(t) = \mathcal{T}_{t}(\rho_{0})$.  The choice of jump operators $\lbrace L_{\mu} \rbrace$ determines the specific RD model.  For the case of the annihilation reaction we have one operator for each pair of nearest neighbours, so that in one dimension there is one operator per site and the label $\mu$ coincides with the site label $i$,
\begin{equation}
    L^{(\text{ann})}_{i} = \sqrt{\kappa} \, \sigma_{i}^{-} \sigma_{i+1}^{-} .
    \label{lann}
\end{equation}
The same occurs for asymmetric coagulation,
\begin{equation}
    L^{(\text{ac})}_{i} = \sqrt{\kappa} \, \sigma_{i}^{+} \sigma_{i}^{-} \sigma_{i+1}^{-} .
        \label{lac}
\end{equation}
For the case of symmetric coagulation there are two jump operators per site since coagulation can be with the left or right neighbour,
\begin{equation}
    L^{(\text{sc})}_{i,\pm} = \sqrt{\kappa/2} \, \sigma_{i}^{+} \sigma_{i}^{-} \sigma_{i \pm 1}^{-} . 
        \label{lsc}
\end{equation}
With these definitions these quantum models are intuitively comparable to their classical counterparts. In particular, as we will see below, the QMEs associated to these three models (and therefore the dynamical properties) are equivalent in a well-defined sense.

\subsection{Invariant subspaces}

One of the main points of interest of this paper is the time evolution of the density of particles. We define the density operator $\Lambda$ to be the global observable $\Lambda := {N^{-1}} \sum_{i=1}^{N} n_{i}$, whose expectation value $\langle \Lambda (t) \rangle = \Tr \left( \Lambda \rho(t) \right)$ is the density of particles.

The operators $H, \Lambda$ and the $\lbrace L_{\mu}^{(\text{m})} \rbrace$ satisfy the algebraic relations
\begin{equation}
    [ H, \Lambda ] = 0, \quad [ L^{\text{(m)}}_{\mu}, \Lambda ] = L^{\text{(m)}}_{\mu}
\label{algebra}
\end{equation}
where the superscript $\text{(m)}$ indicates any of the three models \eqref{lann}-\eqref{lsc} above.  The unravelling of the dynamics in terms of jump operators $\lbrace L_{\mu} \rbrace$ which generates the quantum jump trajectories, is such that the evolution between jumps is governed by the effective Hamiltonian $H_{\text{eff}} := H - \tfrac{i}{2} \sum_{\mu} L_{\mu}^{\dagger} L_{\mu}$. This operator coincides for the three models and takes the form
\begin{equation} 
\label{eq:effham}
    H_{\text{eff}} = H - i \frac{\kappa}{2} \sum_{i} n_{i} n_{i+1} .
\end{equation}
Due to the fact that $\left[ H_{\text{eff}}, \Lambda \right] = 0$, at the level of quantum jump trajectories associated to the models above, the density $\langle \Lambda \rangle$ is a conserved quantity between jumps, only decreasing whenever a jump occurs.  This is the first indication of the dynamical equivalence of the annihilation and coalescence models: both the evolution of the wave-function and the distribution of times between quantum jumps (for example in the quantum jump Monte Carlo unravelling of the dynamics \cite{Plenio1998}) is determined by the $H_{\text{eff}}$ which is the same for the three models.  The only differences are in the post-jump states due to the distinct jump operators. We will return to this equivalence between the three models in the next section. 

Conservation of the particle density $\langle \Lambda \rangle$ between jumps allows us to define a natural decomposition of the system Hilbert space $\mathcal{H}$ into subspaces with a given number of particles (see Fig.\ \ref{fig:subspaces}). We write
\begin{equation}\label{eq:decomposition}
    \mathcal{H} = \bigoplus_{k=0}^{N} \mathcal{H}_{k}
\end{equation}
where the $k$-particle subspace $\mathcal{H}_{k}$ is characterised by $\Tr \left ( \Lambda \rho \right) = k/N$ for all states $\rho \in \mathcal{S}(\mathcal{H}_{k})$; we will later consider this useful decomposition in more detail.

\subsection{Quantum jump trajectories}

\begin{figure}[t]
\centering
    \includegraphics[width=\columnwidth]{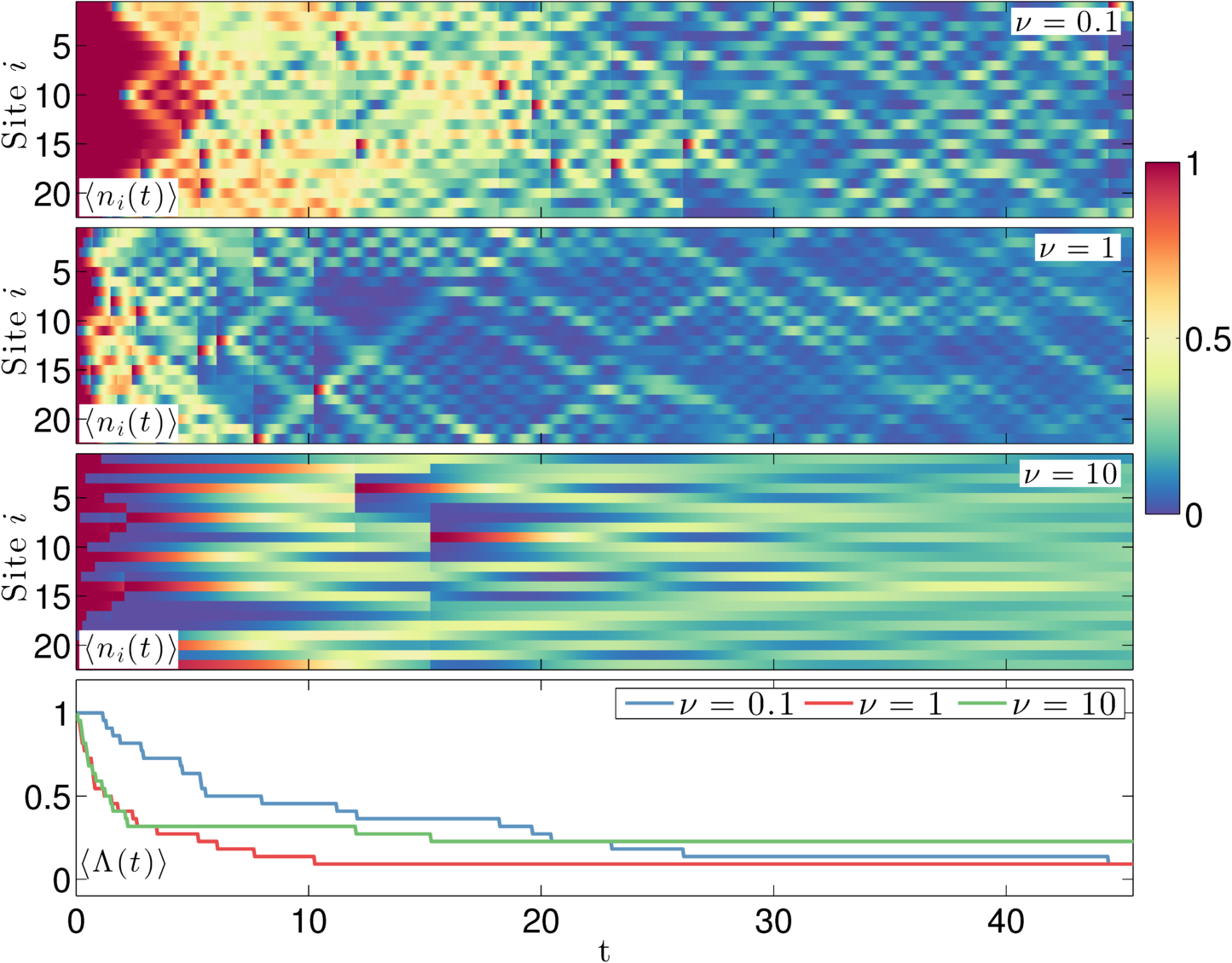}
\caption{Sample quantum jump trajectories associated to the asymmetric coagulation model, for system size $N=22$ sites.  The top three panels show individual site occupation $\langle n_{i} \rangle$ as a function of time.  The bottom panel is the particle density $\langle \Lambda (t) \rangle$.  The ratio $\nu := \kappa / \Omega$ characterises the relative relevance of reactions to diffusion.  In all cases we expect an eventual crossover to a diffusion limited regime at low enough densities, which is most obvious for the case of $\nu \gg 1$: here an initial fast decay which exhibits an almost cutoff-like transition to a much slower relaxation.}
\label{fig:trajectory}
\end{figure}

We follow a quantum Markov chain Monte Carlo approach \cite{Plenio1998} to simulate continuous-time quantum jump trajectories, which are realisations of the unraveling \cite{Wiseman2001} of the master equation in Eq.\ \eqref{eq:masterequation}. In Fig.\ \ref{fig:trajectory} we have plotted the site occupations $\langle n_{i} \rangle$ and the density $\langle \Lambda \rangle$ along such sample trajectories. Two important features, common to all three models, stand out in these trajectories.  The first is that in between quantum jumps the wave function spreads due to the coherent hopping of particles, something that is more evident when the density $\langle \Lambda \rangle$ is lower.  The second feature is the decay of the density towards a stationary value. The rate of decay of $\langle \Lambda (t)\rangle$ is of particular interest to us, and one of our main results in this paper will be the characterisation of the power-law decay of this density.  

We consider quantum jump trajectories with the completely filled initial state $\ket{1, \ldots, 1}$, with initial number of particles $N$ (so $\langle \Lambda \rangle = 1$). The number of particles subsequently decreases whenever a jump occurs, decreasing either by $1$ (in either of the coagulation models) or by $2$ (in the annihilation model). In each trajectory, the number of particles reaches a stationary value; as we will show below, the stationary values available to each trajectory are determined by the existence of stationary states within the sectors $\mathcal{H}_{k}$ from Eq.\ \eqref{eq:decomposition}. We show below that there are such \emph{dark states} in the $k=0$ and $k=1$ sectors, and if $N$ is even, in the $k=2$ sector. The asymptotic values of $\langle \Lambda \rangle$ are, depending on model and parity of $N$, $0, 1/N$ and $2/N$; in the limit of large system size, we of course have $\langle \Lambda \rangle \rightarrow 0$.

We note that in either of the coagulation models the $k=0$ sector is inaccessible when the trajectory is initialised outside of this sector; this is because at least two particles are required for a reaction to occur. The annihilation model conserves parity in the following sense: if $N$ is odd the reachable sector in a quantum jump trajectory is $\mathcal{H}_{1}$, whereas for even $N$ the sectors $\mathcal{H}_{0}$ and $\mathcal{H}_{2}$ are accessible.

A further comment relates to differences in the behaviour of the trajectories as we vary the diffusion rate $\Omega$ relative to the reaction rate $\kappa$.  The classical literature \cite{Hinrichsen2000} distinguishes \emph{diffusion-limited} and \emph{reaction-limited} regimes depending on which of the two processes represents the limiting factor for the evolution of the particle density.  This can refer both to the transient or to the asymptotic time behaviour.  We already see something similar in the quantum jump trajectories of Fig.\ \ref{fig:trajectory}: starting from a fully occupied lattice, in all cases we see that there is a change in behaviour at higher densities where reactions are plentiful, to one at lower densities where particles need to propagate increasingly larger distances for reactions to occur.  If we consider trajectories with different ratios $\nu := \kappa / \Omega$, we see that the crossover between the two regimes occurs earlier for larger $\nu$.  As we will see below, there is a key distinction between the mean-field approximation for the dynamics of these systems and the actual behaviour in one dimension: the former predicts that the asymptotic relaxation is reaction-limited, while the latter is actually hopping-limited, as in the classical problem.  (Aspects of the QME in the limit of low density are also investigated in Ref. \cite{Everest2014}.) Having made these observations, in the remainder of this paper we will set the rates $\kappa = \Omega = 1$.

\subsection{Transient and recurrent subspaces}

\begin{figure}[t]
    \includegraphics[width=0.46\textwidth]{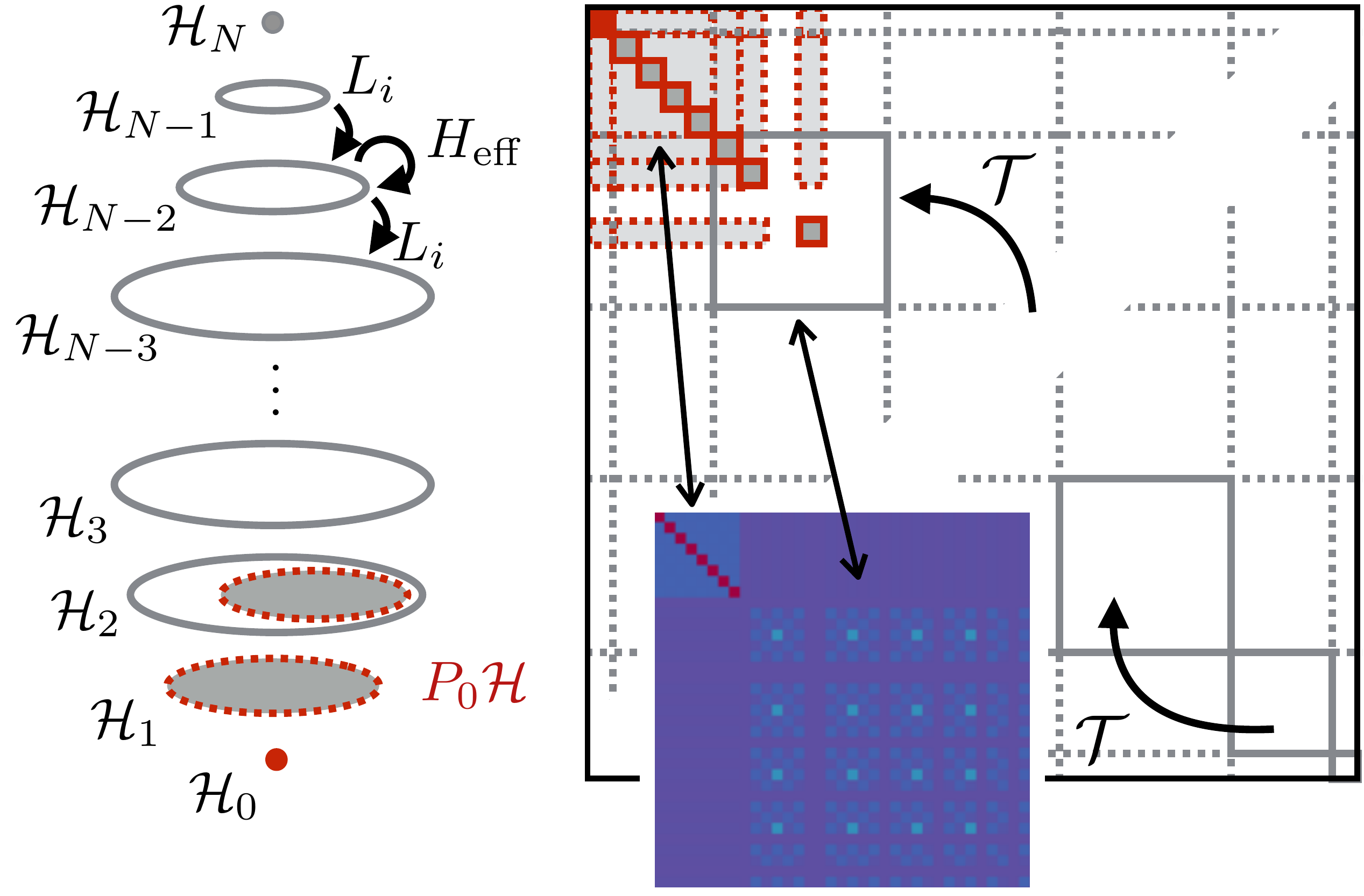}
    \caption{The decomposition of $\mathcal{H}$ in Eq.\ \eqref{eq:decomposition} can be thought of as a cascade of decay for the quantum jump trajectory which ultimately ends up in the recurrent subspace $P_{0} \mathcal{H}$. The corresponding evolution due to the quantum transition operator $\mathcal{T} \equiv \mathcal{T}_{1}$ is sketched on the right; the stationary state will be a mixture of the states in $P_{0} \mathcal{H}$ (dark blocks) and stationary or non-decaying oscillating coherences between them (lighter blocks); shown in inset are absolute values of matrix elements of the actual density operator $\rho(t)$ for large $t$ (obtained by integrating the QME), restricted to the one and two-particle subspaces (for details see Fig.\ \ref{fig:truncatedspectrum}).}\label{fig:subspaces}
\end{figure}

The decomposition of the Hilbert space $\mathcal{H}$ in Eq.\ \eqref{eq:decomposition} may be split up into a direct sum of a transient (or maximal decaying) subspace and a recurrent subspace. As indicated in Fig.\ \ref{fig:subspaces}, along a single quantum trajectory, the state jumps from higher into lower subspaces; collecting together all the subspaces that are eventually empty defines the transient subspace.

Following the terminology of \cite{Baumgartner2008a}, the higher levels $\mathcal{H}_{3} \oplus \ldots \oplus \mathcal{H}_{N}$ can be regarded as a \emph{cascade of decay}, with a natural subdivision into levels $\mathcal{H}_{k} = P_{k} \mathcal{H}$. As the state evolves, it moves strictly down the levels, that is, for any state $\rho \in \mathcal{S}(\mathcal{H})$ and all $t>0$,
\begin{equation*}
    P_{j} \mathcal{T}_{t}\left( P_{k} \rho P_{k} \right) P_{j} = 0 \quad \mbox{for all } j > k.
\end{equation*} 

 Indeed, following \cite[Thm.\ 2]{Baumgartner2008a} we write the system Hilbert space $\mathcal{H}$ as a direct sum $\mathcal{H} = P_{0} \mathcal{H} \oplus P_{0}^{\perp} \mathcal{H}$, where $P_{0}^{\perp} \mathcal{H}$ is the maximal decaying subspace. The latter is defined by requiring that $\lim_{t \rightarrow \infty} P_{0}^{\perp} \mathcal{T}_{t}(\rho) P_{0}^{\perp} = 0$ for all initial states $\rho$, and $P_{0} \mathcal{H}$ contains no further decaying subspace. In other words, the state of the system asymptotically becomes supported on the recurrent subspace $P_{0} \mathcal{H}$.

Since the dissipative part of the dynamics consists of jump operators acting on two adjacent particles, it seems reasonable to suppose that the decaying subspace is given by the Hilbert space of states with two or more particles. However, as we will show below, there are dark states even in the two-particle subspace $\mathcal{H}_{2}$; the recurrent subspace is generated by a fully known set of dark states, and in fact $P_{0} \mathcal{H} \subset \mathcal{H}_{0} \oplus \mathcal{H}_{1} \oplus \mathcal{H}_{2}$ whenever $N$ is even (see Fig.\ \ref{fig:subspaces}).

\begin{figure}
%    %\myfloatalign
%    \subfloat[$N=4$]
%    {\includegraphics[width=0.16\textwidth]{spec0}}
%    \subfloat[$N=5$]
%    {\includegraphics[width=0.24\textwidth]{spec1}}
%    \subfloat[$N=6$]
%    {\includegraphics[width=0.44\textwidth]{spec2}}    
    \includegraphics[width=0.35\textwidth]{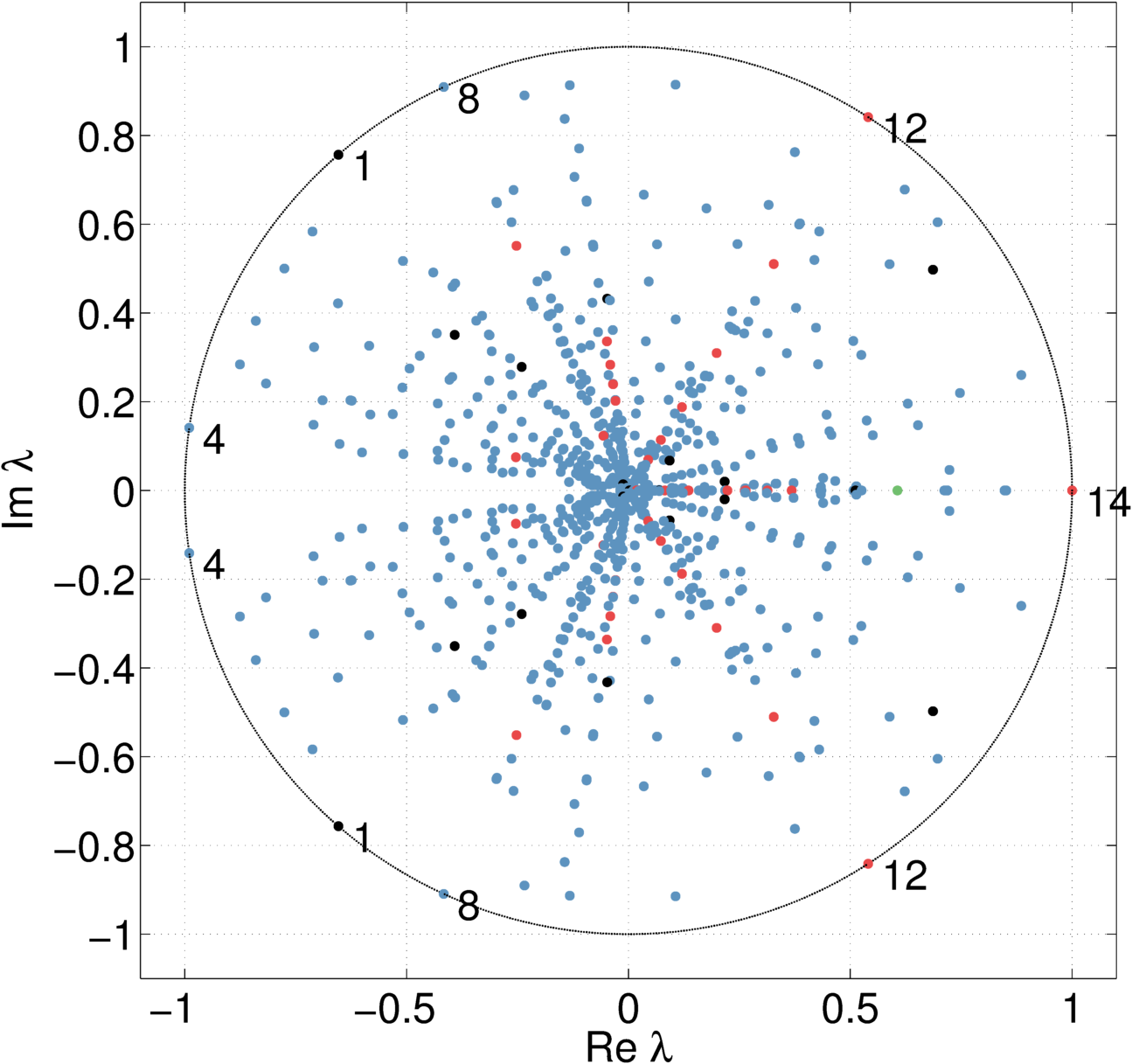}    
    \caption{Spectrum of the transition operator $\mathcal{T}_{1} = \exp \mathbb{W}$ for $N=6$; the numbers indicate the algebraic multiplicities of the peripheral eigenvalues (cf. Fig.\ \ref{fig:truncatedspectrum}).}\label{fig:spec}
\end{figure}

Analysis of the eigenvalues of the master operator $\mathbb{W}$ reveals the structure of the recurrent space beyond the stationary states (see Fig.\ \ref{fig:spec}). The peripheral eigenvalues of $\mathbb{W}$ (i.e. those with vanishing real part) correspond to eigenstates in the recurrent subspace: they are either stationary (if the eigenvalue is $0$) or oscillating without decay. We will show below that each of the peripheral eigenvalues is associated to a specific dark state, thus completely characterising the recurrent subspace.

\section{Dark states and the recurrent subspace}

This section contains our main analytical results, arranged as follows: we first provide a complete classification of the dark states for the class of reaction-diffusion models described in the previous section. We consider the implications of these dark states for the structure of the recurrent subspace $P_{0} \mathcal{H}$ and we end the section with an argument for the equivalence of the three models. If we briefly return our attention to classical RD processes, we note that typically the recurrent subspace is fairly straightforward, consisting of only localised one-particle states which hop indefinitely. As we will see in this section, our quantum models present a more richly structured collection of absorbing states.

\subsection{Dark states}

Recall that a vector $\ket{\psi} \in \mathcal{H}$ is called a \emph{dark state} \cite{Kraus2008} if it is an eigenvector of the Hamiltonian and belongs to the nullspace of all of the jump operators:
\begin{equation}
    H \ket{\psi} = \lambda \ket{\psi}, \;\;\;\;\;
    L_{\mu} \ket{\psi} = 0 \;\;  \forall \mu .\label{dark}
\end{equation}
Note that we may immediate conclude that the dark states are the same for all three models since the nullspace of $L_{\mu}$ is independent of choice of model. We now proceed with one of our main results, which is a classification of the dark states of $\mathbb{W}$ for any of the three models \eqref{lann}-\eqref{lsc}. We treat each of the subspaces separately: it is clear that the singleton subspace $\mathcal{H}_{0} = \set{\ket{0,\ldots,0}}$ is dark. We will now separately consider the subspaces $\mathcal{H}_{1}$ and $\mathcal{H}_{2}$ (where we refer the reader to Appendices A and B for some of the calculations).

\subsubsection{One-particle subspace}

The subspace $\mathcal{H}_{1}$ of states with a single particle is contained in the nullspace of all the jump operators: whenever $\ket{\psi} \in \mathcal{H}_{1}$, we have $L_{\mu}\ket{\psi} = 0$. This means that the task of finding dark states in $\mathcal{H}_{1}$ is reduced to that of diagonalising the restriction of the Hamiltonian $H$ to the subspace $\mathcal{H}_{1}$. The restricted Hamiltonian $H_{1}$ is a translation invariant hopping Hamiltonian, 
\begin{equation}
H_{1} = \left(\begin{array}{cccccc} 0 & 1 & 0 & 0 &  \ldots & 1 \\ 1 & 0 & 1 & 0 & \ldots & 0\\ 0 & 1 & 0 & 1 & \ldots & 0 \\ \vdots & & & & \ddots & \vdots \\ 1 & 0 & 0 & 0 & \ldots & 1 \end{array} \right)\label{eq:ham1}
\end{equation}
in the usual position basis, corresponding to the action $H \ket{k} = \ket{k-1} + \ket{k+1}$. We find (see Appendix A) that the eigenvalues $\set{\lambda_{j}}$ of $H_{1}$ are given by 
\begin{equation} \label{eq:eigsone}
    \lambda_{j} := 2 \cos \left( \frac{2 \pi j}{N} \right),\quad j=0,1,\ldots,n,
\end{equation} 
where $n := \lfloor N/2 \rfloor$. The corresponding eigenvectors are generated by the states
\begin{equation}\label{eq:quasimomentum}
\begin{aligned}
    \ket{\varphi_{k}} &:= \sum_{l=1}^{N} \cos \left( \frac{2 \pi (l-1) k}{N} \right) \ket{l}, \\ \ket{\phi_{k}} &:= \sum_{l=1}^{N} \sin \left( \frac{2 \pi (l-1) k}{N} \right) \ket{l}, \quad k=0,\ldots,n
\end{aligned}
\end{equation}
in the following sense: the largest eigenvalue is $\lambda_{0}=2$ which has the unique eigenvector $\ket{\varphi_{0}} = \sum_{l=1}^{N} \ket{l}$. If $N$ is even, $\lambda_{n} = -2$ is the smallest eigenvalue, and has the unique eigenvector $\ket{\varphi_{n}} = \sum_{l=1}^{N} (-1)^{l} \ket{l}$.

All other eigenvalues are degenerate and have a pair of orthogonal eigenvectors associated to them: for $k=1,\ldots, n-1$ (if $N$ is even) or $k=1,\ldots, n$ (if $N$ is odd), the eigenspace associated to $\lambda_{k}$ is two-dimensional, and is spanned by the vectors $\ket{\varphi_{k}}$ and $\ket{\phi_{k}}$. 

We note the special case when $N$ is a multiple of $4$, in which case the ground states of $H$ in $\mathcal{H}_{1}$ are the eigenvectors associated to the zero eigenvalue $\lambda_{N/4}$, which take the simple form
\begin{equation*}
\begin{aligned}
    \ket{\varphi_{N/4}} &= \ket{1} - \ket{3} + \ldots + \ket{N-3} - \ket{N-1}\\
    \ket{\phi_{N/4}} &= \ket{2} - \ket{4} + \ldots + \ket{N-2} - \ket{N}.
\end{aligned}
\end{equation*}
It is clear from this discussion that $H$ is diagonalisable when restricted to $\mathcal{H}_{1}$; the eigenstates $\ket{\varphi_{j}}$, $\ket{\phi_{j}}$ define an alternative basis for $\mathcal{H}_{1}$, sometimes referred to as the quasi-momentum basis. The subspace $\mathcal{H}_{1}$ is entirely contained in the recurrent subspace $P_{0} \mathcal{H}$; below we will return to the dynamical consequences of this observation.

\subsubsection{Two-particle subspace}

We now classify the dark states in the two-particle subspace $\mathcal{H}_{2}$. We first consider how the basis vectors for $\mathcal{H}_{2}$ are generated; in particular, we express the position basis vectors in terms of the translation operator $T = \sum_{i=1}^{N} \sigma^{-}_{i} \sigma^{+}_{i+1}$ in order to exploit the translation invariance of the Hamiltonian. For example, the basis vectors for $\mathcal{H}_{1}$ are generated by the single vector $\ket{1} = \ket{1,0,\ldots,0} = \sigma^{+}_{1} \ket{0}$ as
\begin{equation*}
    \ket{k} = T^{k-1}\ket{1}.
\end{equation*}
For the two-particle subspace $\mathcal{H}_{2}$, we denote (for $m \geq 2$) by $\ket{\psi_{m}}$ the vector $\ket{1,m} = \sigma^{+}_{1} \sigma^{+}_{m} \ket{0}$. The vectors of this form generate all two-particle basis vectors $\set{\ket{k,m} : k \neq m}$: for $k=1,\ldots,N$,
\begin{equation*}
    \ket{k, k+l-1} = T^{k-1} \ket{\psi_{l}}, \quad l = 2,\ldots, \lfloor N/2 \rfloor + 1.
\end{equation*}
In Appendix B we use this expression for the basis vectors of $\mathcal{H}_{2}$ to derive the equations that any dark state must satisfy. We find that if $N$ is odd, there are no dark states in $\mathcal{H}_{2}$. 

However, when $N$ is even, there are dark states in $\mathcal{H}_{2}$, the number of which increases linearly in $N$. In particular, the dark states are linear combinations of the states
\begin{equation}\label{eq:darkstatestwo1}
    \ket{\Phi_{l}} := \sum_{k=1}^{N} (-1)^{k} T^{k-1} \ket{\psi_{l}},\, l=3,\ldots,N/2
\end{equation}
and, if $n = N/2$ is itself even,
\begin{equation}\label{eq:darkstatestwo2}
    \ket{\Psi_{n+1}} := \sum_{k=1}^{n} (-1)^{k} T^{k-1} \ket{\psi_{n+1}}.
\end{equation}
As we argue in the final section of the Appendix, these are the only dark states in $\mathcal{H}_{2}$. 

In Tables \ref{tab:darkstatesodd} and \ref{tab:darkstateseven} we summarise our classification of dark states, using the notation for the one-particle eigenvalues and quasimomentum states from Eqs.\ \eqref{eq:eigsone} and \eqref{eq:quasimomentum} and the notation for the two-particle dark states from Eqs.\ \eqref{eq:darkstatestwo1} and \eqref{eq:darkstatestwo2}. To argue that there are no dark states to be found in any of the higher sectors, we provide numerical evidence in the remainder of this section.

\begin{table}
\caption{\label{tab:darkstatesodd}Dark states, $N = 2n+1$.}
\begin{ruledtabular}
\begin{tabular}{cccl}
  Subspace & Eigenvectors & Eigenvalue & Dark states\\
%  \hlin
  $ \mathcal{H}_{0}$ & $\ket{0,\ldots,0}$ & $0$  & $ 1$\\
  \ldelim\{{5}{2em}[$\mathcal{H}_{1}$] & $\ket{\varphi_{0}} = \sum_{l=1}^{N} \ket{l}$ & $\lambda_{0} = 2$  & \rdelim\}{5}{2em}[$N$]\\
   & $\ket{\varphi_{1}}, \ket{\phi_{1}}$ & $\lambda_{1}$ &  \\
   & \vdots &  \vdots & \\
   & $\ket{\varphi_{n}}, \ket{\phi_{n}}$ & $\lambda_{n}$ &  \\   
  $ \mathcal{H}_{2}$ & $-$ & $-$ &  $ 0$
\end{tabular}
\end{ruledtabular}

\caption{\label{tab:darkstateseven}Dark states, $N = 2n$.}
\begin{ruledtabular}
\begin{tabular}{cccl}
  Subspace & Eigenvectors & Eigenvalue & Dark states\\
%  \hlin
  $ \mathcal{H}_{0}$ & $\ket{0,\ldots,0}$ & $0$ & $1$ \\
  \ldelim\{{6}{2em}[$\mathcal{H}_{1}$] & $\ket{\varphi_{0}} = \sum_{l=1}^{N} \ket{l}$ & $\lambda_{0} = 2$ & \rdelim\}{6}{2em}[$N$]\\
   & $\ket{\varphi_{1}}, \ket{\phi_{1}}$ & $ \lambda_{1} $ &  \\
   & \vdots & \vdots &\\
   & $\ket{\varphi_{n-1}}, \ket{\phi_{n-1}}$ & $ \lambda_{n-1} $ &  \\   
   & $\ket{\varphi_{n}} = \sum_{l=1}^{N} (-1)^{l} \ket{l}$ & $\lambda_{n} = -2$ & \\
  \ldelim\{{4}{2em}[$\mathcal{H}_{2}$] & $\ket{\Phi_{3}}$ & $0$ &  \rdelim\}{4}{2em}[$2\lfloor N/4 \rfloor - 1$] \\
  & \vdots & \vdots &  \\
  & $\ket{\Phi_{n}}$ & $0$ &  \\
  & $\ket{\Psi_{n+1}}$ ($n$ even) & $0$ &  
\end{tabular}
\end{ruledtabular}
\end{table}

\subsubsection{Numerical analysis}

To further confirm the results summarised in Tables \ref{tab:darkstatesodd} and \ref{tab:darkstateseven} and argue that there are no dark states found in any of the higher subspaces $\mathcal{H}_{k}$, $k \geq 3$, we now discuss related numerical results. We approach the problem in two ways: we perform an exhaustive search for dark states within each of the subspaces $\mathcal{H}_{k}$, and we exactly diagonalise the master operator, accounting for all the non-decaying eigenvalues solely using the dark states in Tables \ref{tab:darkstatesodd} and \ref{tab:darkstateseven}.

\begin{figure}
    %\myfloatalign
    \includegraphics[width=.4\textwidth]{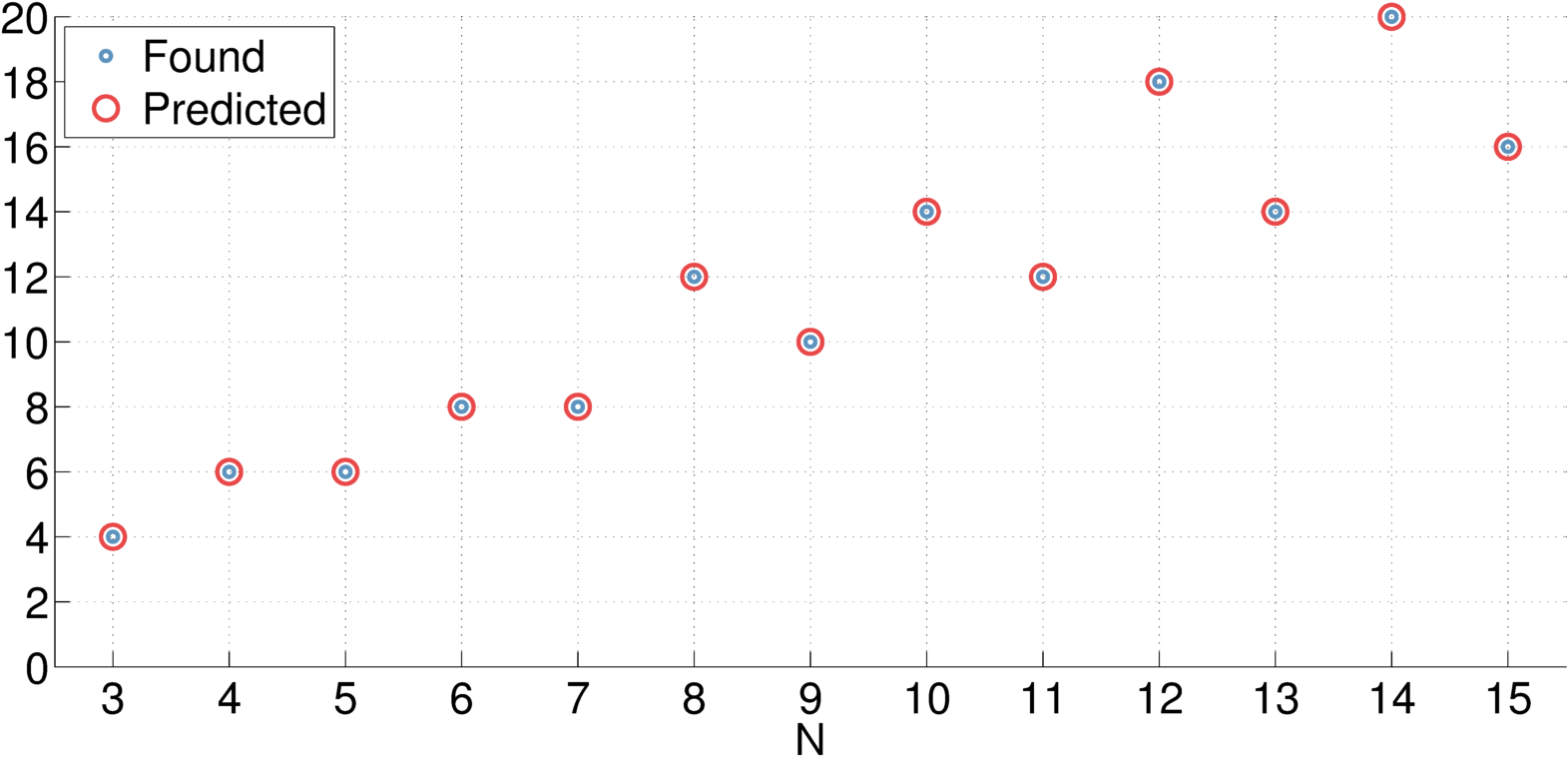}              
    \caption{Number of dark states on a chain of $N$ sites predicted in Tables \ref{tab:darkstatesodd} and \ref{tab:darkstateseven} by our analytical results, and found by numerical search.}\label{fig:darkstatesearch}
\end{figure}

We search for dark states $\ket{\varphi}$ satisfying \eqref{dark} in each of the subspaces $\mathcal{H}_{k}$.  That is $\ket{\varphi} \in \mathcal{H}_{k}$ with
\begin{equation}
    H_{k} \ket{\psi} = \lambda \ket{\psi}, \;\;\;\;\;
    L_{\mu,k} \ket{\psi} = 0 \;\;  \forall \mu ,
\end{equation}
where $H_{k}$ and $L_{\mu,k}$ are the operators restricted to $\mathcal{H}_{k}$. We fully diagonalise $H_{k}$ and for each eigenvalue $\ket{\lambda_{i}}$ of $H_{k}$, we look for a linear combination of the associated eigenvectors $\set{\ket{\varphi_{1}^{i}},\ldots, \ket{\varphi_{n}^{i}}}$ which is in the nullspace of all the jump operators. This means solving the system of linear equations with matrix of coefficients
\begin{equation*}
    \mathcal{C}_{k} =  
    \left(\begin{array}{ccc} L{1,k} \ket{\varphi^{i}_{1}} & \cdots & L_{1,k} \ket{\varphi^{i}_{n}} \\ \vdots & \ddots & \vdots \\ L_{M,k} \ket{\varphi^{i}_{1}} & \cdots & L_{M,k} \ket{\varphi^{i}_{n}}\end{array} \right) ,
\end{equation*}
where $N'$ is the total number of jump operators ($N'=N$ for the pair-annihilation and asymmetric coagulation processes, $N'=2N$ for the symmetric coagulation process).  The number of linearly independent dark states in $\mathcal{H}_{k}$ is then given by the number of linearly independent solutions this system of equations, which is given by $n - \textrm{rank}( \mathcal{C}_{k})$. This quantity is plotted in Fig.\ \ref{fig:darkstatesearch} for various $N$, showing full agreement with the predicted number of dark states from Tables \ref{tab:darkstatesodd} and \ref{tab:darkstateseven}. 

For further confirmation of our results, and to clarify their role in the asymptotic behaviour of the dynamics, we consider the spectral properties of the master operator. Mixtures of dark states are eigenstates of $\mathbb{W}$ corresponding to its \emph{peripheral} eigenvalues -- that is, eigenvalues $\lambda$ of $\mathbb{W}$ which are either $0$ or purely imaginary. To make this point clearer, suppose $\ket{\varphi}$ and $\ket{\psi}$ are two dark states for $\mathbb{W}$ corresponding to eigenvalues $\lambda$ and $\mu$ of $H$, respectively.
%\begin{equation*}
%    H \ket{\varphi} = \lambda \ket{\varphi},\quad H \ket{\psi} = \mu \ket{\psi}.
%\end{equation*}
Then $\mathbb{W}(\ket{\varphi}\bra{\psi}) = -i(\lambda - \mu ) \ket{\varphi}\bra{\psi}$, that is, $\ket{\varphi}\bra{\psi}$ is an eigenmatrix of $\mathbb{W}$ corresponding to the peripheral eigenvalue $-i(\lambda - \mu )$: in general, if there are $M$ dark states, the sum of the algebraic multiplicities of the peripheral eigenvalues \emph{associated to dark states} of $\mathbb{W}$ must be $M^{2}$. This, of course, does not rule out the existence of other stationary states of $\mathbb{W}$ which are not pure; however, we argue along numerical lines that the peripheral eigenvalues are entirely accounted for by the eigenvalues obtained in this way from dark states.

\begin{figure}
    %\myfloatalign
    \includegraphics[width=0.34\textwidth]{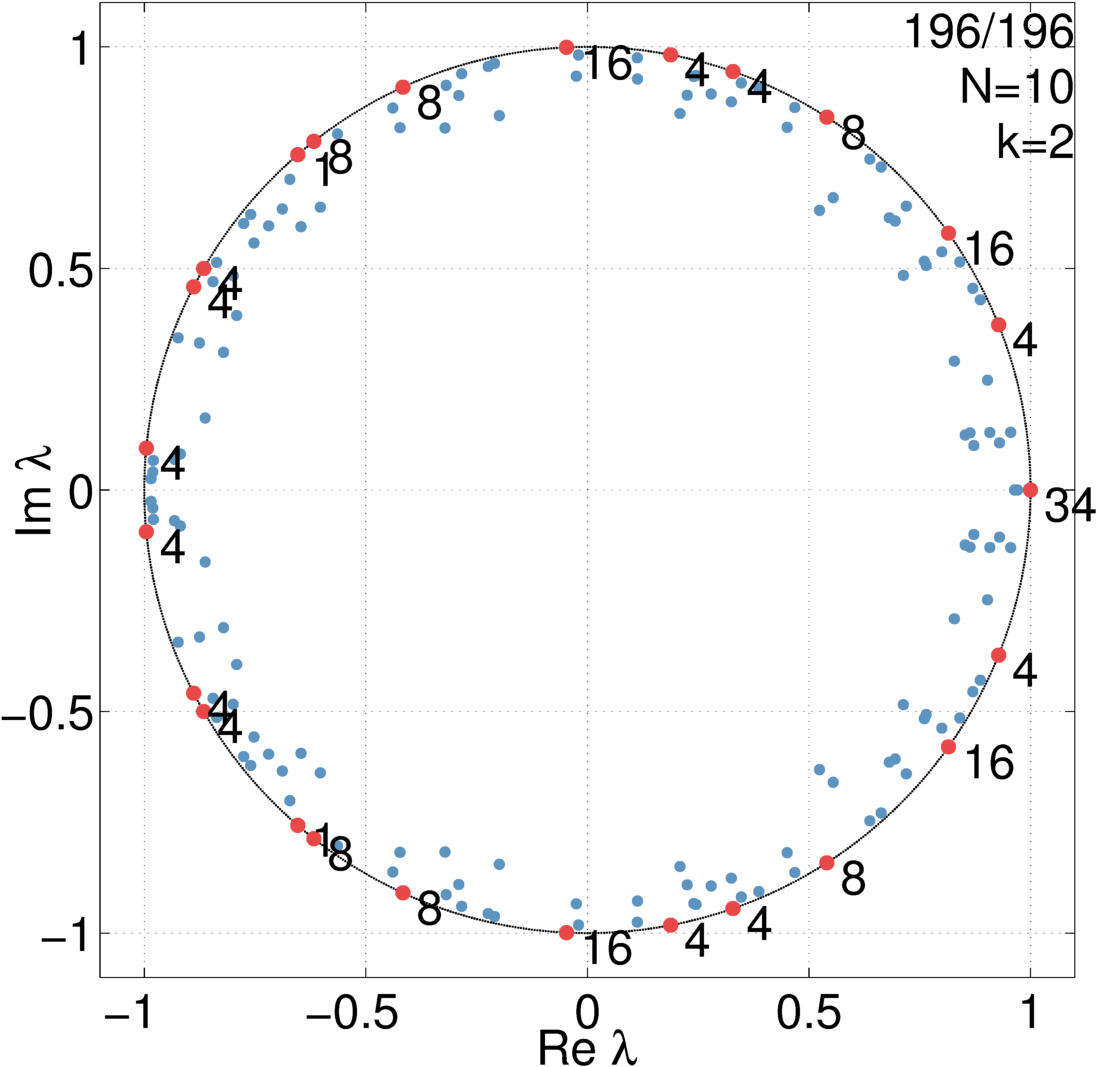}                  
    \caption{Partial spectrum of $\mathcal{T}_{1} = \exp \mathbb{W}$ restricted to $k=2$ lowest sectors with $N=10$; as predicted the total algebraic multiplicity of the peripheral eigenvalues is $14^{2} = 196$. Each of the peripheral eigenvalues is of the form $e^{i (\lambda_{i} - \lambda_{j})}$ where $\lambda_{i}$ and $\lambda_{j}$ are eigenvalues of $H$ associated to the dark states found in Tables \ref{tab:darkstatesodd} and \ref{tab:darkstateseven}. }\label{fig:truncatedspectrum}
\end{figure}

Although, computationally speaking, full diagonalisation of $\mathbb{W}$ for larger $N$ becomes prohibitively expensive, we are able to extract the peripheral eigenvalues by restricting the master operator to the lowest $k$ particle subspaces. We thus obtain a list of peripheral eigenvalues $\lambda_{1}, \ldots, \lambda_{m}$ of $\mathbb{W}^{(k)}$ along with their multiplicities; for example, for $N=10$, the results are shown in Fig.\ \ref{fig:truncatedspectrum}. We find that, for values of $N$ up to $10$, the total algebraic multiplicity of the peripheral eigenvalues is indeed $M^{2}$, where $M$ is the number of dark states. Furthermore, if $\tau$ is one of the resulting peripheral eigenvalues of $\mathbb{W}$, there is a pair of eigenvalues $\lambda, \mu$ associated to dark states such that $\tau = -i(\lambda - \mu )$.

This confirms that the peripheral eigenvalues are exactly determined by the dark states listed in Tables \ref{tab:darkstatesodd} and \ref{tab:darkstateseven}. Conversely, all peripheral eigenvalues are accounted for by exact diagonalisation in the sectors up to two particles; extending the diagonalisation scheme to higher sectors results in eigenvalues of $\mathbb{W}$ with larger negative real part, as expected from the cascade of decay discussed in Fig.\ \ref{fig:subspaces}.

We briefly expand on the structure of the recurrent subspace $P_{0} \mathcal{H}$. If $N$ is odd, the recurrent subspace consists only of $\mathcal{H}_{0} \oplus \mathcal{H}_{1}$, and the asymptotic dynamics is purely unitary on this subspace: as $t \rightarrow \infty$ we have $\mathcal{T}_{t}(\rho) \rightarrow U_{t}^{\dagger} \rho U_{t}$ where $U_{t} = \exp (i H t).$ This remains true when $N$ is even, but the recurrent subspace $P_{0} \mathcal{H}$ has an additional component in the subspace $\mathcal{H}_{2}$ spanned by the two-particle dark states in Table \ref{tab:darkstateseven}.

\begin{figure}
    %%\myfloatalign
    \includegraphics[width=0.4\textwidth]{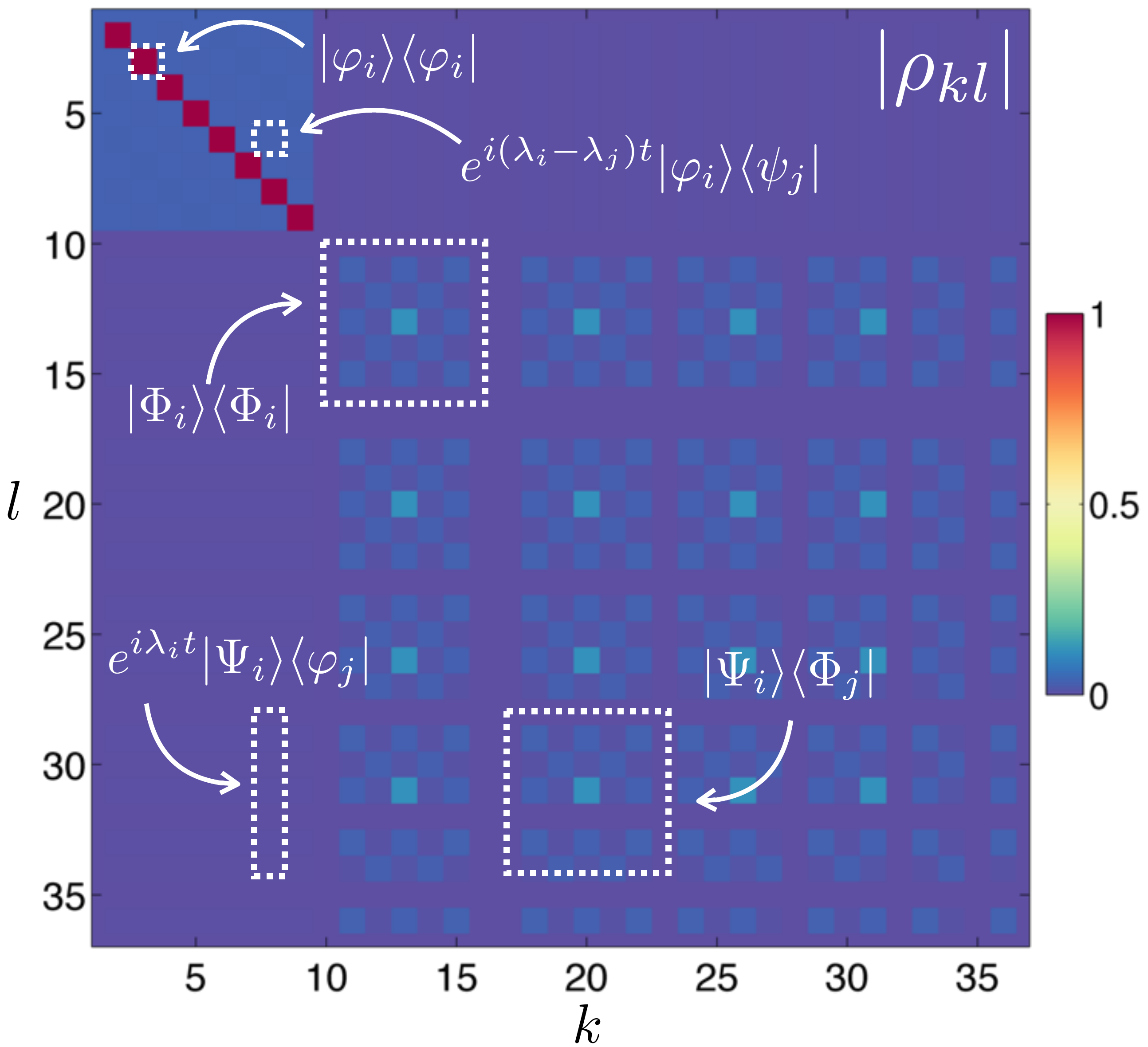}                  
    \caption{Structure of the recurrent subspace $P_{0} \mathcal{H}$ appearing in the matrix elements of  the asymptotic state $\rho(t)$ obtained by numerical integration of the master equation for $N=8$. The non-vanishing entries in $\rho(t)$ are generated by the dark states found in Table \ref{tab:darkstateseven}, which are indicated in this figure. The off-diagonal blocks are coherences between dark states, which oscillate at a rate determined by the difference of the corresponding eigenvalues (\emph{Note}: the highlighted regions are only for intuitive purposes; the actual dark states are spread out in the position basis).}\label{fig:recurrentstructure}
\end{figure}

In either case, dark states corresponding to the same eigenvalue of $H$ lead to pure stationary states for $\mathbb{W}$; dark states corresponding to different eigenvalues of $H$ lead to oscillating coherences in the state $\rho(t)$, appearing in the off-diagonal blocks. In Fig.\ \ref{fig:recurrentstructure} we show how the matrix elements of the state $\rho(t)$ for large $t$ reflects this structure. Note that these observations about the structure of $P_{0} \mathcal{H}$ serve to illustrate the general theory concerning the asymptotic structure of the time evolution of quantum dynamical semigroups found in e.g.\ Refs.\ \cite{Baumgartner2008a,Albert2014}. 

%\begin{figure}
%    %\myfloatalign
%%    \subfloat[]
%%    {\includegraphics[width=.3\textwidth]{darkstatesfound15}}\\
%    \subfloat[$N=7, k = 2$]
%    {\includegraphics[width=0.2\textwidth]{truncatedspectrum-7-2}}
%    \subfloat[$N=8, k = 2$]
%    {\includegraphics[width=0.2\textwidth]{truncatedspectrum-8-2}}\\
%    \subfloat[$N=9, k = 2$]
%    {\includegraphics[width=0.2\textwidth]{truncatedspectrum-9-2}}
%    \subfloat[$N=10, k = 2$]
%    {\includegraphics[width=0.2\textwidth]{truncatedspectrum-10-2}}                   
%    \caption{Spectrum of $\mathcal{T}_{1} = \exp \mathbb{W}$ restricted to $k$ lowest sectors; as predicted, for $N=7,8,9$ and $10$, with $k \geq 2$, the total number of peripheral eigenvalues is $64,144,100$ and $196$, respectively.}\label{fig:truncatedspectrum}
%\end{figure}

\subsection{Reducibility of the dynamics and reachability of dark states}

Our results concerning the dark states for this class of reaction-diffusion models allow us to conclude that the recurrent subspace is generated by the set of dark states. The question remains if it is possible to reach any state in the recurrent subspace by choosing an appropriate initial state -- a process of interest in quantum control \cite{Mabuchi2005}, where it is known as quantum state engineering \cite{Verstraete&Wolf&Cirac09}. Aside from this, the question also relates to what occurs in the classical pair-annihilation model \cite{Hinrichsen2000} where there is a conservation of parity, based on which we can always predict whether the stationary state will be $0$ or a $1$-particle state.

\begin{figure}[]
    %\myfloatalign
    \includegraphics[width=.5\textwidth]{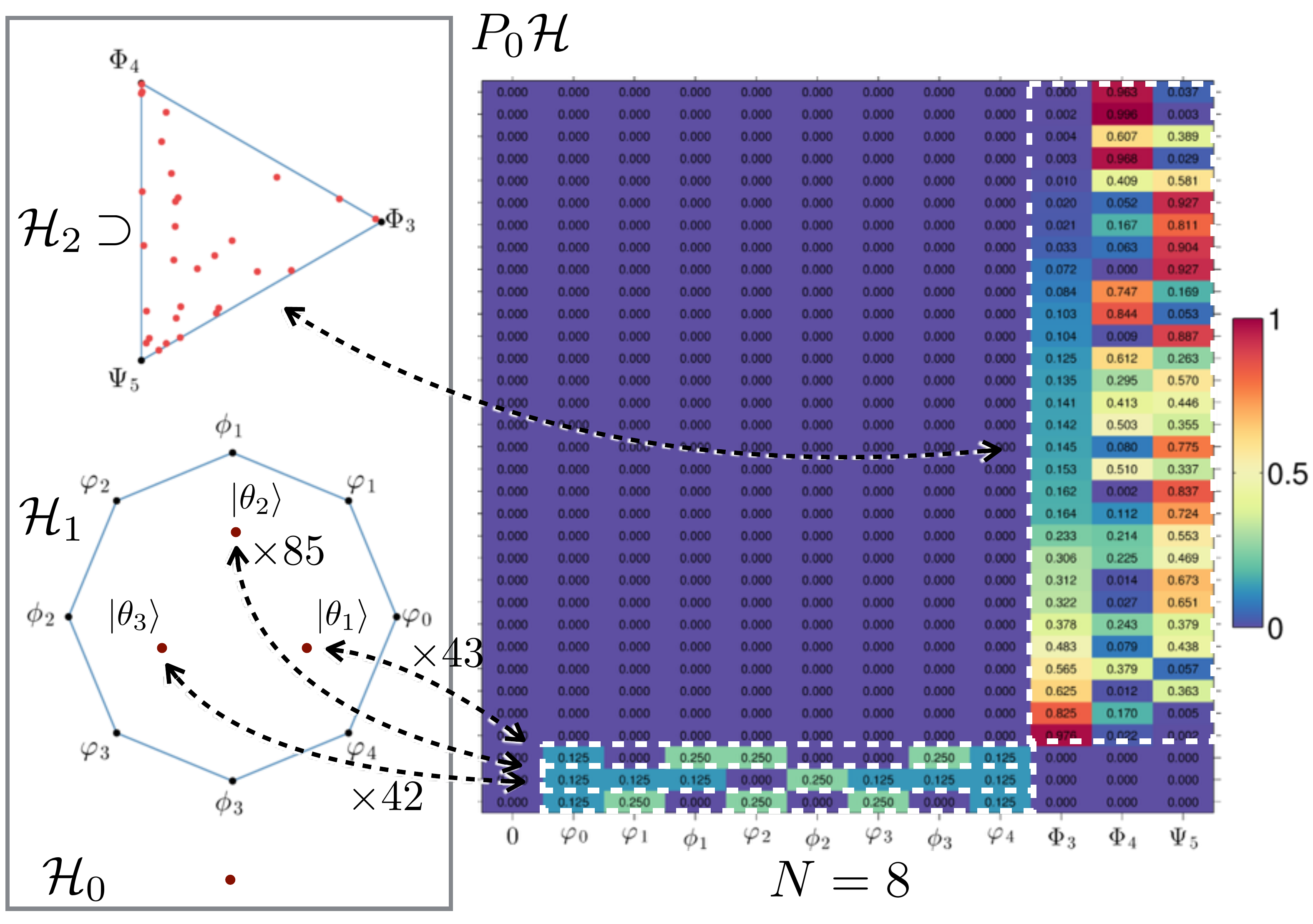}   
%    \subfloat[]
%    {\includegraphics[width=.4\textwidth]{fuguestates-n8-full}}
    \caption{Asymptotic behaviour of quantum jump trajectories in terms of dark states, obtained by taking final states of 200 trajectories and collecting identical states. The one- and two-particle dark subspaces are represented as convex sets with the dark states in Tables \ref{tab:darkstatesodd} and \ref{tab:darkstateseven} as their extremal points. The reachable one-particle states $\ket{\theta_{i}}$ are indicated in the interior of $\mathcal{H}_{1}$; the reachable two-particle states appear randomly distributed in the interior of the two-particle dark subspace.}\label{fig:darkstate}
\end{figure}

Although a full treatment of engineering of dark states is beyond the scope of this paper, we will briefly consider \emph{reachability} \cite{Ying2013} of the dark states we have previously identified. In the current context, the question of reachability is as follows: given a state in the recurrent subspace $\theta \in \mathcal{S}\left(P_{0}\mathcal{H}\right)$ (which is necessarily a mixture of dark states), is it possible to find an initial state $\theta_{0}$ in the \emph{transient} subspace such that $\mathcal{T}_{t}(\theta_{0}) \rightarrow \theta$ as $t \rightarrow \infty$? 

Our results indicate a negative answer to this question; that is, it appears that for \emph{pure} states starting outside of the recurrent subspace $P_{0} \mathcal{H}$, only a few superpositions of the one-particle dark states are available as stationary states. We approach this problem as follows: starting with the full initial state, we compute an ensemble of trajectories, each of which until it has reached stationarity -- that is, until the quantum state is fully supported on $P_{0} \mathcal{H}$. We then take each final state and compute its inner product with each of the dark states found in Tables \ref{tab:darkstatesodd} and \ref{tab:darkstateseven}. Recalling that the dark states form an orthogonal basis of $P_{0} \mathcal{H}$, these inner products uniquely determine the final trajectory states.

We have plotted the magnitudes of the resulting coefficients in Fig.\ \ref{fig:darkstate}, for the case $N=8$, taking an ensemble of $200$ trajectories. It appears that, whenever the initial state does \emph{not} have any dark components, the final state state component in the one-particle space is limited to one of only a few possible states, while the component in the two-particle space is chosen randomly.  Rephrasing in terms of reachability, the few one-particle dark states that appear asymptotically are reachable, while the others are not -- their coefficients need to be present initially. In the case of Fig.\ \ref{fig:darkstate} with $N=8$, the reachable states in $\mathcal{H}_{1}$ are
\begin{equation*}
\begin{aligned}
    \ket{\theta_{1}} &:= \tfrac{1}{8} \left( \ket{\varphi_{0}} + 2 \ket{\phi_{1}} + 2 \ket{\varphi_{2}} + \ket{\phi_{3}} + \ket{\varphi_{4}} \right),\\
    \ket{\theta_{2}} &:= \tfrac{1}{8} \left( \ket{\varphi_{0}} +  \ket{\varphi_{1}} + \ket{\phi_{1}} + 2 \ket{\phi_{2}} + \ket{\varphi_{3}} + \ket{\phi_{3}} + \ket{\varphi_{4}} \right),\\
    \ket{\theta_{3}} &:= \tfrac{1}{8} \left( \ket{\varphi_{0}} + 2 \ket{\varphi_{1}} + 2 \ket{\varphi_{2}} + 2 \ket{\varphi_{3}} + \ket{\varphi_{4}} \right).
\end{aligned}
\end{equation*}
As indicated in Fig.\ \ref{fig:darkstate}, out of the $200$ final states, these occur $43$, $85$ and $49$ times, respectively; the remaining states are randomly distributed elements of the two-particle dark space.

Repeating this experiment with a randomly chosen initial state vector (outside of $P_{0}\mathcal{H}$) confirms these results: the only states reached in $\mathcal{H}_{1}$ are the $\ket{\theta_{i}}, i=1,2,3$. Based on this, we conjecture that not all dark states are reachable from a state which starts outside of the dark space. We conclude our observations on reachable states by noting that the coefficients which appear in the vectors $\theta_{i}$ (for general $N$) are exactly the coefficients of the dark states in Eq.\ \eqref{eq:quasimomentum}; this is certainly an interesting property of the reachable states, and the question remains open whether this is an indication of some deeper symmetry in the dynamics.

\subsection{Equivalence of models}

\begin{figure}[t]
    %\myfloatalign
    \includegraphics[width=0.6\columnwidth]{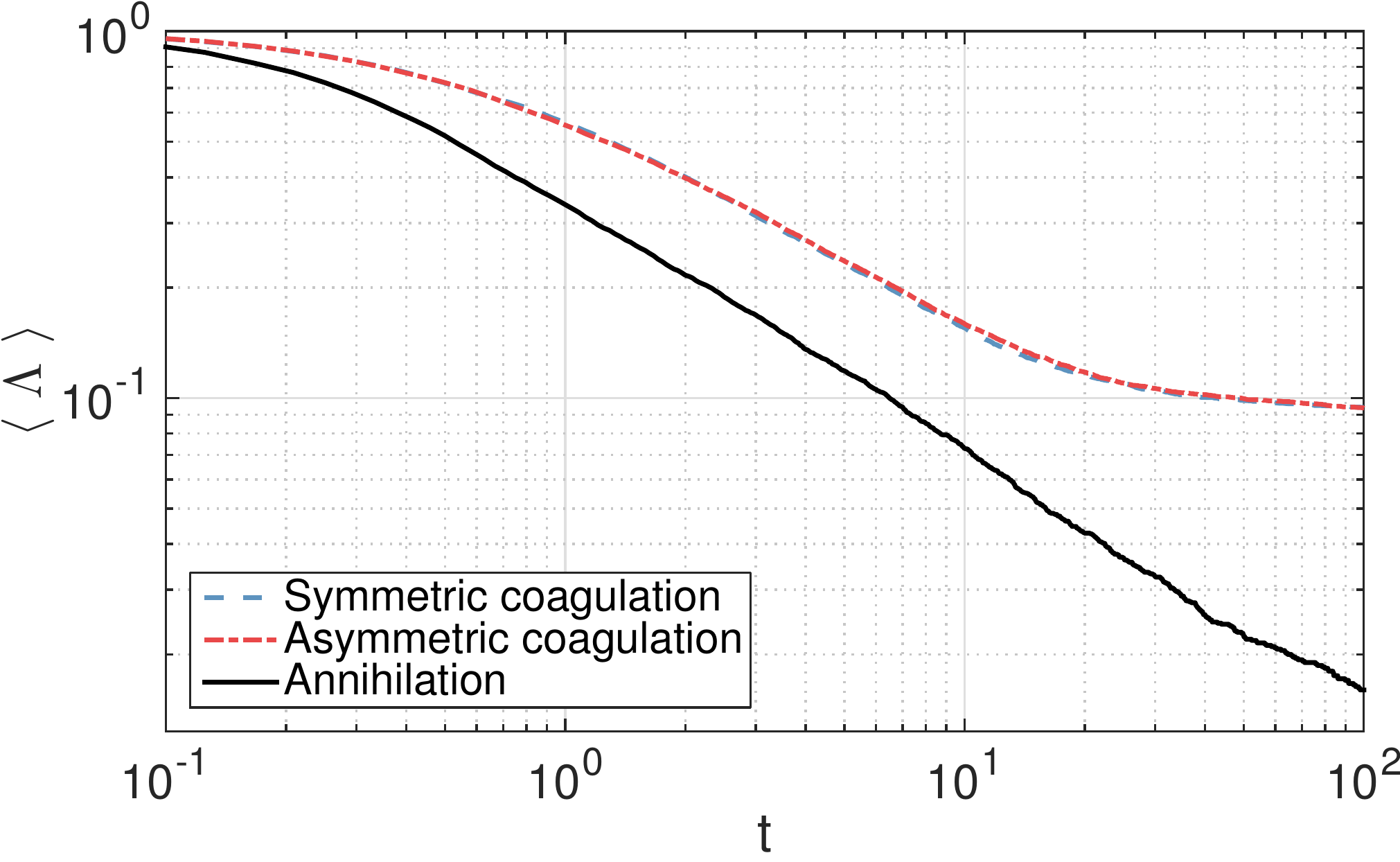}                    
    \caption{Comparison of density $\langle \Lambda(t) \rangle$ from quantum jump trajectory averages for the three models for $N=12$. The coagulation models show a non-zero final density, while the density for the annihilation model decays to zero.}\label{fig:compthree}
\end{figure}

The classical reaction-diffusion processes have been shown to be equivalent in the sense that their Liouvillian operators are related by similarity transformations \cite{Hinrichsen2000}.  This notion of equivalence may be extended to quantum systems by letting two quantum systems, defined by their master operators (or quantum Liouvillians), be equivalent if their master operators are related by a similarity transformation \cite{Burgarth2012,Burgarth2014}. This relation, in turn, may be defined via the Choi-Jamiolkowski isomorphism on the level of the operator-sum representations of the master operators, by requiring that their matrix representations are similar in the usual sense. That is, since the operator-sum representation is an isomorphism, the notion of matrix similarity between the matrix representations of linear maps immediately defines similarity of master operators: if $\mathbb{W}$ and $\tilde{\mathbb{W}}$ are two master operators we may say that $\mathbb{W}$ and $\tilde{\mathbb{W}}$ are \emph{similar} if there exists an invertible linear map $\mathcal{P}$ such that $\tilde{\mathbb{W}} = \mathcal{P}^{-1} \circ \mathbb{W} \circ \mathcal{P}$. 

This allows us to call the quantum systems defined by master operators equivalent \emph{if} their master operators are similar in this well-defined sense. Expected properties for equivalent systems are immediate from this definition: in particular, two master operators have equivalent if and only if they have exactly the same spectrum. If this is the case, the eigenstates are related by the similarity transformation.

\begin{figure*}[t]
    \includegraphics[width=0.92\textwidth]{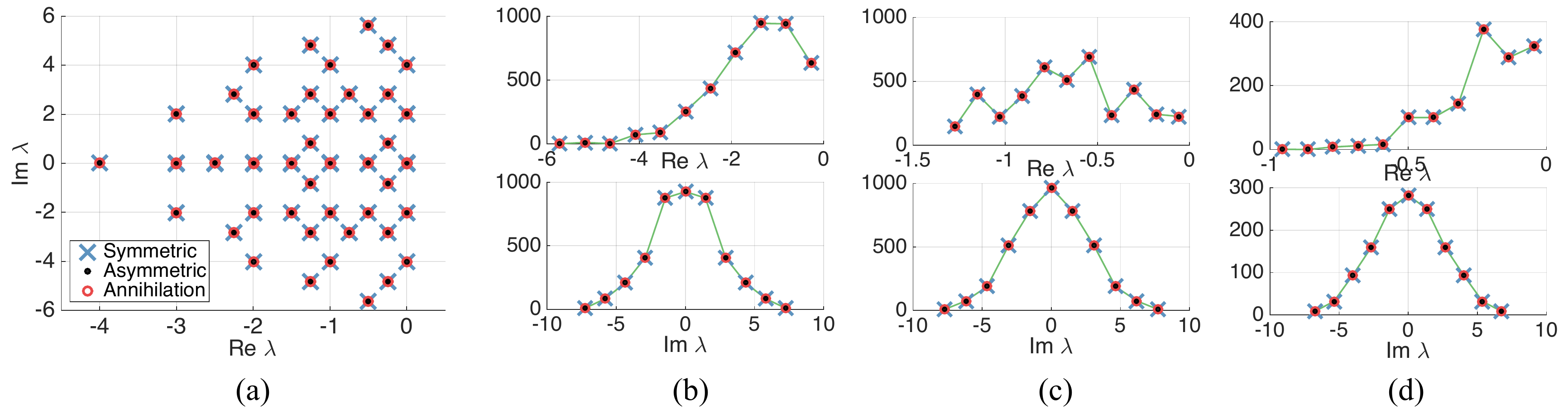}                    
    \caption{(a) Comparison of the spectrum of $\mathbb{W}$ for the three models, with overlapping symbols showing complete agreement ($N=4$). (b) For clarity, instead of plotting all eigenvalues for larger $N$ we show the density of states as function of the real and imaginary part of the spectrum ($N=6$); (c) $N=7$, truncated to three particles; (d) $N=8$, truncated to two particles.}\label{fig:equiv}
\end{figure*}

For the three master operators $\mathbb{W}^{\text{(ann)}}$, $\mathbb{W}^{\text{(ac)}}$ and $\mathbb{W}^{\text{(sc)}}$ defined in Eq.\ \eqref{eq:masterequation}-\eqref{lsc} we have plotted the density $\langle \Lambda(t) \rangle$ in Fig.\ \ref{fig:compthree}. On the level of the expectation value of observables such as $\Lambda$, a similarity transformation between the master operators would need to likewise transform between the densities $\langle \Lambda(t) \rangle$. For evidence that the master operators $\mathbb{W}^{\text{(m)}}$ are indeed equivalent, we have computed their eigenvalues and compared them for increasing system size; for $N=6$, for example, see Fig.\ \ref{fig:equiv}. Indeed, it appears that, as far as we are able to numerically diagonalise the master operators, the eigenvalues (and their algebraic multiplicities) agree, and the models are indeed equivalent in the sense discussed above.

\section{Decay of particle density}

Having established the asymptotic behaviour of the three equivalent quantum reaction-diffusion models by classifying the dark states and the structure of the recurrent space, we now turn to the dynamics of the system and consider how the density $\langle \Lambda(t) \rangle$ decays. We start by working out the mean-field approximation of the site densities $\langle n_{l}(t) \rangle$, which results in a power-law prediction for the density of the form $\langle \Lambda(t)\rangle \sim t^{-1}$. We will then perform an analysis of the actual behaviour of the density as obtained from simulations of the dynamics; it turns out that a power law decay still holds, but with a slower rate than predicted by the mean-field approximation.

\subsection{Mean-field approximation}

Using the Gutzwiller product state ansatz $\rho = \bigotimes_{i} \rho_{i}$ we approximate the equations of motion for the expectation values of observables $n_{l}, \sigma^{+}_{l}$ and $\sigma^{-}_{l}$:
\begin{align*}
    \partial_{t} \langle n_{l} \rangle &= i \Omega \left( \langle \sigma_{l-1}^{+} \rangle + \langle \sigma_{l+1}^{+} \rangle \right) \langle \sigma_{l}^{-} \rangle\\ -& i \Omega \left( \langle \sigma_{l-1}^{-} \rangle + \langle \sigma_{l+1}^{-} \rangle \right) \langle \sigma_{l}^{+} \rangle - \tfrac{1}{2} \kappa \langle n_{l-1} \rangle \langle n_{l} ,\rangle\\
    \partial_{t} \langle \sigma_{l}^{+} \rangle &= -i \Omega \left( \langle \sigma_{l-1}^{+} \rangle + \langle \sigma_{l+1}^{+} \rangle \right) \langle \sigma_{l}^{z} \rangle\\ &- \tfrac{1}{2} \kappa \left( \langle n_{l-1} \rangle + \langle n_{l+1} \rangle \right) \langle \sigma_{l}^{+} \rangle,\\
        \partial_{t} \langle \sigma_{l}^{-} \rangle &= i \Omega \left( \langle \sigma_{l-1}^{-} \rangle + \langle \sigma_{l+1}^{-} \rangle \right) \langle \sigma_{l}^{z} \rangle\\ &- \tfrac{1}{2} \kappa \left( \langle n_{l-1} \rangle + \langle n_{l+1} \rangle \right) \langle \sigma_{l}^{-} \rangle,
\end{align*}
where $\sigma^{z} = 2n - \mathbf{1}$. Assuming a homogeneous state, so that one-site expectation values are all equal, we arrive at the mean-field equation of motion for the average density $\kappa \langle n \rangle$,
\begin{equation*}
    \partial_{t} \langle n \rangle = -\frac{1}{2} \kappa \langle n \rangle^{2}.
\end{equation*}
This equation is readily solved to obtain a power-law decay with exponent $-1$,
\begin{equation} \label{eq:sitemeanfield}
    \langle n (t) \rangle = \left( \frac{1}{2} \kappa t + \nu_{0}^{-1} \right)^{-1} \sim (\kappa t)^{-1},
\end{equation}
where $\nu_{0}$ denotes the initial mean site density. This is the same type of behaviour obtained in the mean-field predictions for the analogous classical reaction-diffusion processes \cite{Hinrichsen2000}.  Note that the mean-field result also predicts that the asymptotic regime is reaction-limited as it only depends on $\kappa$.  In the classical case, however, the mean-field law is only accurate above two dimensions \cite{Hinrichsen2000,Tauber2005}: in one dimension spatial fluctuations dominate, the asymptotic regime is diffusion-limited, and the density decays as $1/\sqrt{D t}$.  A natural question is whether an analogous departure from mean-field behaviour is observed in our one-dimensional quantum models.

\subsection{Numerics in one dimension}

\begin{figure*}[t]
    \includegraphics[width=0.92\textwidth]{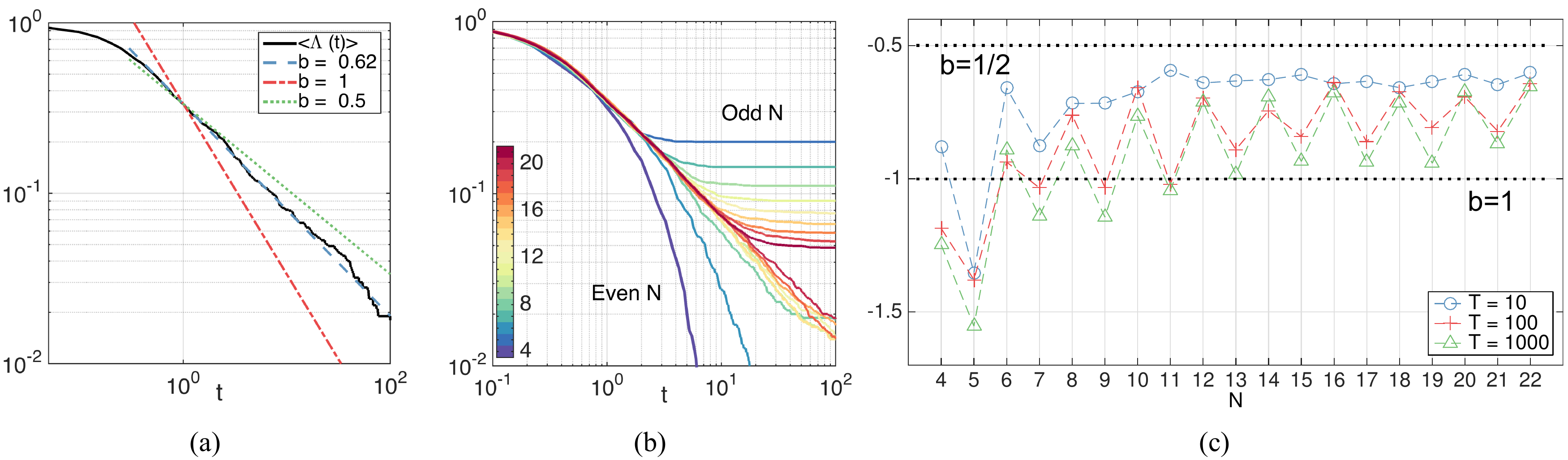}   
    \caption{(a) Density $\langle \Lambda(t) \rangle$ for $N=20$ from quantum jump Monte Carlo, compared to power law fit. (b) Convergence of density $\langle \Lambda(t) \rangle$ for $N = 4, \ldots, 20$. (c) Convergence of the power law exponent (the plot shows $-b$) with increasing system size $N$ (dependence on trajectory length $T$ (arb. units) is also indicated.)}
    \label{fig:densitydecay}
\end{figure*}

From the quantum jump simulations we obtain the time evolution of the particle density $\langle \Lambda(t) \rangle$.  As noted in the previous section, the three models defined in Eq.\ \eqref{eq:masterequation} are equivalent; in particular, if a power law decay holds for the density in one of the models, it is true for all of them. We therefore mainly consider the pair-annihilation model in the following discussion, unless otherwise specified. 

In Fig.\ \ref{fig:densitydecay}(a) we have plotted the density $\langle \Lambda (t) \rangle$ for $N=20$ sites, obtained by taking an average of $100$ trajectories, on a log-log scale. After an initial transient period, the density seems to follow a power law.  If we fit a function of the form $a \, t^{-b} + c$ to this data, we obtain $\langle \Lambda(t) \rangle \approx 0.34 \, t^{-0.62}$, with a vanishing coefficient $c$, which is expected in the pair-annihilation model for even $N$ and a fully occupied initial state: the two-particle dark states in the subspace are highly unlikely to be visited, and hence any trajectory almost always ends up in the zero-particle state. If $N$ is odd, the trajectory will always converge to the one-particle subspace, resulting in a non-vanishing coefficient $c$. 

This distinction between odd and even $N$ is clearly visible in Fig.\ \ref{fig:densitydecay}(b), where we have plotted the trajectory average density $\langle \Lambda(t) \rangle$ for $N$ ranging from $4$ to $20$. The even-numbered sites decay to $0$ while the odd-numbered sites approach a non-zero asymptotic value determining the coefficient $c$. However, since the density computed on the subspace $\mathcal{H}_{1}$ is $N^{-1}$, in the limit of large chain lengths the difference between odd and even $N$ vanishes.  

In order to estimate the decay exponent $b$ we have analysed fits as the one shown in Fig.\ \ref{fig:densitydecay}(a) for varying trajectory lengths for times between $10$ and $1000$, for system sizes up to $N=22$. As shown in Fig.\ \ref{fig:densitydecay}(c), the distinct final state reached for even or odd sites impinges on the fitted value of $b$,  
with even $N$ giving exponents $b \gtrsim 0.6$ and odd $N$ giving exponents $b \lesssim 0.9$. This disparity between odd and even $N$ fits appears to diminish as the number of sites increases.  Extrapolation to large $N$ of our 
simulations would suggest a coefficient $b$ in the range $0.9 < b < 0.7$. As mentioned previously, the difference between the asymptotic values for $\langle \Lambda(t)\rangle$, computed for odd and even $N$, should vanish based on the dynamics alone. A better indication of the actual behaviour in this large system limit, however, is beyond our computational abilities due to the usual exponential-in-size complexity of quantum dynamics.

Our numerical results suggest that dynamics of the quantum RD models studied here are fluctuation dominated in one dimension, and therefore disagree with a mean-field prediction, in analogy to what occurs in the classical case.  In particular, the decay exponent $b$ is smaller than the mean-field prediction, $b < b_{\rm mf} = 1$.  The numerical results, albeit for small systems, suggest however that $b$ is larger than the one-dimensional classical exponent, $b > b_{\rm class.} = 1/2$.  One can speculate that this is due to the fact that in the quantum case particle propagation is via a quantum random walk \cite{Aharonov1993} which explores space somewhat more efficiently than classical diffusion.  This potential discrepancy between the classical and quantum decay exponents is an interesting point that would require more solid confirmation.

\section{Conclusions and outlook}

Our aim in this paper was to understand which features, characteristic of single-species classical reaction-diffusion systems, remain when particle propagation becomes coherent.  The open quantum systems we considered were natural generalisations of the classical $A+A \to \varnothing$ and $A+A \to A$ reaction-diffusion models.  We have found that many of the features of the classical systems are also present in the quantum models.  As expected the quantum systems were shown to display a slow relaxation towards absorbing states.  In particular, we were able to provide a thorough classification of the recurrent (i.e.\, absorbing) subspace in terms of the dark states of the quantum master operator that generates the dynamics.  These absorbing subspaces have a much richer structure in the quantum models than in the classical ones, something that may be of potential interest from the point of view of quantum state preparation. 
We have also provided strong evidence for the dynamical equivalence of the annihilation and coalescence quantum models, in analogy with the equivalence of the corresponding classical models.  This was done through the coincidence of the spectra of the dynamical super-operators from direct numerical diagonalisation of finite systems.  It would be interesting to find a  similarity transformation that maps the quantum models, to bring this equivalence to an exact footing as in the classical case.  Our final result is the observation that in one dimension the relaxation of the density seems to follow a power law with an exponent which is smaller than that of the mean-field analysis.  This is again analogous to what occurs in the classical models.  In the quantum case, however, our numerics would suggest that the exponent is not the same as the classical one, which would be an indication of a distinctive quantum feature in the non-equilibrium dynamics.  This interesting possibility deserves further investigation.

\begin{acknowledgments}
We thank E. Levi, M. Hush and J. Schick for helpful discussions.  This work was supported by EPSRC grant no.\ EP/L50502X/1 and grant no.\ EP/J009776/1.

\end{acknowledgments}

\bibliography{library}

\appendix
\section{One-particle dark states}
In this Appendix we derive the eigenvectors and eigenvalues of the Hamiltonian $H_{1}$ restricted to the one-particle subspace $\mathcal{H}_{1}$, which, since the entire subspace $\mathcal{H}_{1}$ is in the nullspace for all of the jump operators, are also dark states for the master operator $\mathbb{W}$. 

Starting with the restricted Hamiltonian $H_{1}$ in Eq.\ \eqref{eq:ham1}, we note that it takes a particular form, which we exploit to characterise its eigenvectors. Recall that a complex $N \times N$ matrix $A$ is called a \emph{circulant} matrix \cite{Gray2005} if it is of the form
\begin{equation*}
A = \left(\begin{array}{ccccc} c_{0} & c_{1} & c_{2} & \cdots & c_{N-1} \\ c_{N-1} & c_{0} & c_{1} & \cdots & c_{N-2} \\  c_{N-2} & c_{N-1} & c_{0} & \cdots & c_{N-3} \\ \vdots & \vdots & \vdots & \ddots & \vdots \\ c_{1} & c_{2} &  c_{3} & \cdots & c_{0} \end{array} \right),
\end{equation*}    
a matrix completely specified by its initial row $(c_{0}, c_{1}, \ldots, c_{N-1})$. It is easy to see that  $H_{1}$ is a circulant matrix with initial row $(0,1,0,\ldots,0,1)$. The eigenvalues of the general circulant matrix $A$ are \cite{Gray2005} 
\begin{equation*}
    \lambda_{j} = \sum_{k=0}^{N-1} c_{k} \omega_{j}^{k}, \quad j=0,1,\ldots,N-1
\end{equation*}
with corresponding eigenvectors 
\begin{equation*}
    v_{j} = (1, \omega_{j}, \omega_{j}^{2}, \ldots, \omega_{j}^{N-1})',
\end{equation*}
where for $j=0,1,\ldots, n-1$ we denote by $\omega_{j} \equiv \exp ( 2 \pi i j / N )$ the $N$-th roots of unity. Therefore, the eigenvalues of $H_{1}$ are given by
%\begin{equation*}
%\begin{aligned}
%    \lambda_{j} &= \omega_{j} + \omega_{j}^{N-1}\\
%        &= \omega_{j} + \omega_{j}^{-1}\\
%        &= \exp \left( \tfrac{ 2 \pi i j}{ N} \right) + \exp \left( - \tfrac{ 2 \pi i j }{ N} \right)\\
%        &= 2 \cos \left( \tfrac{2 \pi j}{N} \right);
%\end{aligned}
%\end{equation*} 
\begin{equation*}
    \lambda_{j} = \omega_{j} + \omega_{j}^{N-1} = 2 \cos \left( \tfrac{2 \pi j}{N} \right);
\end{equation*}        
note that $\lambda_{j} = \lambda_{N-j}$ for all $0 < j < N/2$, resulting in $\lfloor N/2 \rfloor + 1$ distinct eigenvalues.

For the eigenvectors of $H_{1}$, we start with the eigenvalue $\lambda_{0} = 2$, which has a unique corresponding eigenvector $v_{0} = (1, \ldots, 1)'$. For each pair of coinciding eigenvalues $\lambda_{j} = \lambda_{N-j}$, with $0 < j < N/2$, there exists a two-dimensional eigenspace generated by the vectors $v_{j}, v_{N-j}$.

%\begin{equation*}
%%\begin{aligned}
%    \omega_{j}^{m} - \omega_{N-j}^{m} = \exp \left( \tfrac{2 \pi i m j}{N}\right) - \exp \left( \tfrac{2 \pi i m (N-j)}{N} \right) = 2 i \sin \left( \tfrac{2 \pi m j}{N} \right)
%%\end{aligned}
%\end{equation*}
We note that for $1 \leq m \leq N-1$ we have the identities $\omega_{j}^{m} - \omega_{N-j}^{m} = 2 i \sin \left( \tfrac{2 \pi m j}{N} \right)$ and similarly $\omega_{j}^{m} + \omega_{N-j}^{m} = 2 \cos \left( \tfrac{2 \pi m j}{N} \right)$.  By taking the difference $v_{j} - v_{N-j}$ and the sum $v_{j} + v_{N-j}$ we obtain an alternative pair of vectors $\varphi_{j}, \phi_{j}$ spanning the eigenspace,
\begin{equation*}
\begin{aligned}
    \varphi_{j} &= (1, \cos \left( \tfrac{2 \pi j}{N} \right), \ldots, \cos \left( \tfrac{2 \pi j (N-1)}{N} \right))',\\ \phi_{j} &= (0, \sin \left( \tfrac{2 \pi j}{N} \right), \ldots, \sin \left( \tfrac{2 \pi j (N-1)}{N} \right))'.
\end{aligned}
\end{equation*}
We finally note that if $N$ is even, the eigenvalue $\lambda_{N/2}$ is simple with unique eigenvector $(1, -1, \ldots, 1, -1)'$. The identification $(1, 0, \ldots, 0) = \ket{1}, \ldots, (0, \ldots, 0, 1) = \ket{N}$ results in the expression $\ket{\varphi_{k}}, \ket{\phi_{k}}$ for the eigenvectors in Eq.\ \eqref{eq:quasimomentum}. 
%Alternatively, the vectors $v_{j}$ and $v_{N-j}$ correspond to the representation
%\begin{equation*}
%\begin{aligned}
%    \ket{\hat{\varphi}_{k}} &= \sum_{l=1}^{N} \exp \left( \frac{2 \pi (l-1) k i}{N} \right) \ket{l},\\ \quad \ket{\hat{\phi}_{k}} &= \sum_{l=1}^{N} \exp \left( - \frac{2 \pi (l-1) k i}{N} \right) \ket{l}, \quad k=0,\ldots,n.
%\end{aligned}
%\end{equation*}

\section{Two-particle dark states}

In this Appendix we derive the dark states in the two-particle subspace $\mathcal{H}_{2}$. We start by finding a convenient notation for the basis vectors for $\mathcal{H}_{2}$ in terms of the translation operator $T$. As discussed in the main text, the vectors of the form $\ket{\psi_{l}} := \ket{1,l} = \sigma^{+}_{1} \sigma^{+}_{l} \ket{0}$ generate the basis vectors of $\mathcal{H}_{2}$. 

To see how the basis vectors $\ket{k,k'}$ are generated, first suppose $N$ is odd, say $N = 2n+1$. Then there are $n$ generating vectors $\ket{\psi_{l}}$: for $l = 2,\ldots,n+1$ we have
\begin{equation*}
    \ket{k,k+l-1} = T^{k-1} \ket{\psi_{l}} \quad \left( T^{N} \ket{\psi_{l}} = \ket{\psi_{l}} \right)
\end{equation*}
for $k=1,\ldots,N$, generating the $n \cdot N = {N \choose 2}$ basis vectors of $\mathcal{H}_{2}$. 

Now suppose $N$ is even, say $N = 2n$. Then there are $n$ generating vectors $\ket{\psi_{l}}$: for $l = 2, \ldots, n$ we have
\begin{equation*}
    \ket{k,k+l-1} = T^{k-1} \ket{\psi_{l}}\quad \left( T^{N} \ket{\psi_{l}} = \ket{\psi_{l}} \right)
\end{equation*}
for $k=1,\ldots,N$ and with $l=n+1$
\begin{equation*}    
    \ket{k,k+n} = T^{k-1} \ket{\psi_{n+1}} \quad \left( T^{n} \ket{\psi_{n+1}} = \ket{\psi_{n+1}} \right)
\end{equation*}
for $k=1,\ldots,n$, generating the $(n-1)\cdot N + N/2 = {N \choose 2}$ basis vectors of $\mathcal{H}_{2}$. 

%For example, with $N=6$ the three generating vectors are $\ket{\psi_{2}} = \ket{1,2}$, $\ket{\psi_{3}} = \ket{1,3}$ and $\ket{\psi_{4}} = \ket{1,4}$ and these generate the basis vectors as follows:
%\begin{equation*}
%\begin{aligned}
%    \ket{1,1,0,0,0,0} &= \ket{\psi_{2}}, \quad &\ket{1,0,1,0,0,0} &= \ket{\psi_{3}}, \quad &\ket{1,0,0,1,0,0} &= \ket{\psi_{4}},\\
%    \ket{0,1,1,0,0,0} &= T \ket{\psi_{2}}, \quad &\ket{0,1,0,1,0,0} &= T \ket{\psi_{3}}, \quad &\ket{0,1,0,0,1,0} &= T \ket{\psi_{4}},\\
%    \ket{0,0,1,1,0,0} &= T^{2} \ket{\psi_{2}}, \quad &\ket{0,0,1,0,1,0} &= T^{2} \ket{\psi_{3}}, \quad &\ket{0,0,1,0,0,1} &= T^{2} \ket{\psi_{4}},\\
%    \ket{0,0,0,1,1,0} &= T^{3} \ket{\psi_{2}}, \quad &\ket{0,0,0,1,0,1} &= T^{3} \ket{\psi_{3}},\\
%    \ket{0,0,0,0,1,1} &= T^{4} \ket{\psi_{2}}, \quad &\ket{1,0,0,0,1,0} &= T^{4} \ket{\psi_{3}},\\
%    \ket{1,0,0,0,0,1} &= T^{5} \ket{\psi_{2}}, \quad &\ket{0,1,0,0,0,1} &= T^{5} \ket{\psi_{3}}.\\
%\end{aligned}
%\end{equation*}
Note that the basis vectors generated by $\ket{\psi_{2}}$ are exactly the pure states of the form $\ket{k,k+1}$ with two adjacent particles. Since $\mathcal{H}_{2} \cap \mathrm{ker} L_{k}$ consists only of scalar multiples of $\ket{k,k+1}$, dark states in $\mathcal{H}_{2}$ are the eigenvectors of $H_{2}$ which are superpositions of all basis states \emph{except} those generated by $\ket{\psi_{2}}$. The remainder of this Appendix concerns the characterisation of these eigenvectors; we consider separately the cases when $N$ is odd and when $N$ is even. As we are only considering elements of $\mathcal{H}_{2}$ in the following, we will omit the subscript on $H_{2}$.

\subsection{Odd $N$}

%%%%%% (written notes Delta 13 - 15)

Let $N$ be odd with $N = 2n+1$ and let $\ket{\Psi} \in \mathcal{H}_{2}$; then using the position basis $\set{T^{k-1} \ket{\psi_{l}}}$ introduced above we may write
\begin{equation*}
    \ket{\Psi} = \sum_{l=2}^{n+1} \sum_{k=1}^{N} \lambda_{k}^{(l)} T^{k-1} \ket{\psi_{l}}
\end{equation*}
where $\lambda_{k}^{(l)} = \langle k, k+l-1 \vert \Psi \rangle$. To determine if $\ket{\Psi}$ is an eigenvector of $H$ we first consider the action of $H$ on the basis vectors: we find that $H \ket{\psi_{l}}$ is given by
%\begin{equation} \label{eq:oddaction}
%    H \ket{\psi_{l}} = \begin{cases} (T^{-1} + 1) \ket{\psi_{3}} &\mbox{if } l = 2,\\
%    (T^{-1} +1) \ket{\psi_{l+1}} + (T+1) \ket{\psi_{l-1}}  &\mbox{if } 3 \leq l \leq n,\\
%    (T^{-1} +1) T^{n+1} \ket{\psi_{n+1}} + (T+1) \ket{\psi_{n}} &\mbox{if } l = n+1. \end{cases}
%\end{equation}
\begin{equation} \label{eq:oddaction}
\begin{aligned}
    (T^{-1} + 1) \ket{\psi_{3}} &\mbox{ if } l = 2,\\
    (T^{-1} +1) \ket{\psi_{l+1}} + (T+1) \ket{\psi_{l-1}}  &\mbox{ if } 3 \leq l \leq n,\\
    (T^{-1} +1) T^{n+1} \ket{\psi_{n+1}} + (T+1) \ket{\psi_{n}} &\mbox{ if } l = n+1.
\end{aligned}    
\end{equation}

Using translation invariance of $H$ we obtain the expression 
%\begin{equation*}
%\begin{aligned}
%    H \ket{\Psi} &= \sum_{l=2}^{n+1} \sum_{k=1}^{N} \lambda_{k}^{(l)} T^{k-1} H \ket{\psi_{l}}\\
%        &= \sum_{k=1}^{N} \lambda_{k}^{(2)} T^{k-1} H \ket{\psi_{2}} + \sum_{l=3}^{n} \sum_{k=1}^{N} \lambda_{k}^{(l)} T^{k-1} H \ket{\psi_{l}} + \sum_{k=1}^{N} \lambda_{k}^{(n+1)} T^{k-1} H \ket{\psi_{n+1}}
%\end{aligned}
%\end{equation*}
\begin{equation*}
\begin{aligned}
    &H \ket{\Psi} = \sum_{k=1}^{N} \lambda_{k}^{(2)} T^{k-1} H \ket{\psi_{2}}\\ &+ \sum_{l=3}^{n} \sum_{k=1}^{N} \lambda_{k}^{(l)} T^{k-1} H \ket{\psi_{l}} + \sum_{k=1}^{N} \lambda_{k}^{(n+1)} T^{k-1} H \ket{\psi_{n+1}};
\end{aligned}
\end{equation*}
for the first term, we use Eq.\ \eqref{eq:oddaction} and, by relabelling indices and using the periodicity condition $\lambda^{(l)}_{k+N} := \lambda^{(l)}_{k}$, find
\begin{equation*}
    \sum_{k=1}^{N} \lambda_{k}^{(2)} T^{k-1} H \ket{\psi_{2}} = \sum_{k=1}^{N} ( \lambda_{k+1}^{(2)} + \lambda_{k}^{(2)} ) T^{k-1} \ket{\psi_{3}}.
\end{equation*}    
Similarly, the second and third terms are given by
\begin{widetext}
\begin{equation*}
\begin{aligned}
\sum_{l=3}^{n} \sum_{k=1}^{N} \lambda_{k}^{(l)} T^{k-1} H \ket{\psi_{l}} &= \sum_{k=1}^{N} ( \lambda_{k-1}^{(3)} + \lambda_{k}^{(3)}) T^{k-1} \ket{\psi_{2}} + \sum_{k=1}^{N} ( \lambda_{k-1}^{(4)} + \lambda_{k}^{(4)})T^{k-1} \ket{\psi_{3}}\\
     &+ \sum_{l=4}^{n-1} \sum_{k=1}^{N} ( \lambda_{k+1}^{(l-1)} + \lambda_{k}^{(l-1)} + \lambda_{k}^{(l+1)} + \lambda_{k-1}^{(l+1)} ) T^{k-1} \ket{\psi_{l}}\\
     &+ \sum_{k=1}^{N} ( \lambda_{k+1}^{(n-1)} + \lambda_{k}^{n-1} ) T^{k-1} \ket{\psi_{n}} + \sum_{k=1}^{N} ( \lambda_{k+1}^{(n)} + \lambda_{k}^{(n)} )T^{k-1} \ket{\psi_{n+1}}
\end{aligned}
\end{equation*}
and
\begin{equation*}
    \sum_{k=1}^{N} \lambda_{k}^{(n+1)} T^{k-1} H \ket{\psi_{n+1}} = \sum_{k=1}^{N} ( \lambda_{k-1}^{(n+1)} + \lambda_{k}^{(n+1)}) T^{k-1} \ket{\psi_{n}} + \sum_{k=1}^{N} ( \lambda_{k-n}^{(n+1)} + \lambda_{k-n-1}^{(n+1)}) T^{k-1} \ket{\psi_{n+1}}
\end{equation*}
\end{widetext}
respectively. We thus arrive at the expression
\begin{equation*}
\begin{aligned}
    &H \ket{\Psi} = \sum_{k=1}^{N} ( \lambda_{k-1}^{(3)} + \lambda_{k}^{(3)}) T^{k-1} \ket{\psi_{2}} \\
    &+ \sum_{l=3}^{n} \sum_{k=1}^{N} ( \lambda_{k+1}^{(l-1)} + \lambda_{k}^{(l-1)} + \lambda_{k}^{(l+1)} + \lambda_{k-1}^{(l+1)} ) T^{k-1} \ket{\psi_{l}}\\
    &+ \sum_{k=1}^{N} ( \lambda_{k+1}^{(n)} + \lambda_{k}^{(n)} + \lambda_{k-n}^{(n+1)} + \lambda_{k-n-1}^{(n+1)} )T^{k-1} \ket{\psi_{n+1}}.
    \end{aligned}
\end{equation*}
Equating coefficients in the eigenvalue equation $H \ket{\Psi} = c \ket{\Psi}$, we obtain the system of equations for $k=1,\ldots,N$
\begin{equation*}
\begin{aligned}
    \lambda_{k-1}^{(3)} + \lambda_{k}^{(3)} &= c \lambda_{k}^{(2)},\\
    \lambda_{k+1}^{(l-1)} + \lambda_{k}^{(l-1)} + \lambda_{k}^{(l+1)} + \lambda_{k-1}^{(l+1)} &= c \lambda_{k}^{(l)},\; l=3,\ldots, n,\\
    \lambda_{k+1}^{(n)} + \lambda_{k}^{(n)} + \lambda_{k-n}^{(n+1)} + \lambda_{k-n-1}^{(n+1)} &= c \lambda_{k}^{(n+1)}.
\end{aligned}
\end{equation*}
As noted previously in this Appendix, for dark states we additionally require that $\lambda_{k}^{(2)} = 0$ for all $k$. Then $\lambda^{(3)}_{k-1} + \lambda^{(3)}_{k} = 0$ for all $k$; using periodicity and the fact that $N$ is odd, this means that $\lambda^{(3)}_{k} = 0$ for all $k$. Using the above system of equations, this implies that all $\lambda^{(4)}_{k}$ vanish, which means that all $\lambda^{(5)}_{k}$ vanish, and so on. We conclude that there are no dark states in $\mathcal{H}_{2}$ when $N$ is odd.

\subsection{Even $N$}

%%%%% Delta 16 - 19

Now suppose $N$ is even with $N=2n$, and let $\ket{\Psi} \in \mathcal{H}_{2}$ have the decomposition into basis vectors
\begin{equation*}
    \ket{\Psi} = \sum_{l=2}^{n} \sum_{k=1}^{N} \lambda_{k}^{(l)} T^{k-1} \ket{\psi_{l}} +  \sum_{k=1}^{n} \lambda_{k}^{(n+1)} T^{k-1} \ket{\psi_{n+1}}
\end{equation*}
where we have isolated the $l = n+1$ term since the corresponding coefficients are periodic in $n$ rather than $N$. As in the case of $N$ odd, we first consider the action of $H$ on the generating vectors $\ket{\psi_{l}}$; in this case $H \ket{\psi_{l}}$ is given by
%\begin{equation} \label{eq:evenaction}
%    H \ket{\psi_{l}} = \begin{cases} (T^{-1} + 1) \ket{\psi_{3}} &\mbox{if } l = 2,\\
%    (T^{-1} +1) \ket{\psi_{l+1}} + (T+1) \ket{\psi_{l-1}}  &\mbox{if } 3 \leq l \leq n,\\
%    (T^{n} + 1)(T+1) \ket{\psi_{n}} &\mbox{if } l = n+1. \end{cases}
%\end{equation}
\begin{equation} \label{eq:evenaction}
\begin{aligned}
    (T^{-1} + 1) \ket{\psi_{3}} &\mbox{ if } l = 2,\\
    (T^{-1} +1) \ket{\psi_{l+1}} + (T+1) \ket{\psi_{l-1}}  &\mbox{ if } 3 \leq l \leq n,\\
    (T^{n} + 1)(T+1) \ket{\psi_{n}} &\mbox{ if } l = n+1.
\end{aligned}
\end{equation}
As in the previous case, we use these expressions to find a suitable expression for $H \ket{\Psi}$; we find that
\begin{equation*}
\begin{aligned}
    &H \ket{\Psi} = \sum_{k=1}^{N} ( \lambda_{k-1}^{(3)} + \lambda_{k}^{(3)}) T^{k-1} \ket{\psi_{2}} \\
    &+ \sum_{l=3}^{n} \sum_{k=1}^{N} ( \lambda_{k+1}^{(l-1)} + \lambda_{k}^{(l-1)} + \lambda_{k}^{(l+1)} + \lambda_{k-1}^{(l+1)} ) T^{k-1} \ket{\psi_{l}}\\
    &+ \sum_{k=1}^{n} ( \lambda_{k+1}^{(n)} + \lambda_{k}^{(n)} + \lambda_{k+n+1}^{(n)} + \lambda_{k+n}^{(n)} )T^{k-1} \ket{\psi_{n+1}}.
    \end{aligned}
\end{equation*}
Using this expression and equating coefficients in $H \ket{\Psi} = c \ket{\Psi}$ we obtain the following eigenvalue equations for even $N$: for $k=1,\ldots,N$
\begin{equation*}
\begin{aligned}
    \lambda_{k-1}^{(3)} + \lambda_{k}^{(3)} &= c \lambda_{k}^{(2)},\\
    \lambda_{k+1}^{(l-1)} + \lambda_{k}^{(l-1)} + \lambda_{k}^{(l+1)} + \lambda_{k-1}^{(l+1)} &= c \lambda_{k}^{(l)},\; l=3,\ldots, n,
\end{aligned}
\end{equation*}
and for $k=1,\ldots,n$
\begin{equation*}
    \lambda_{k+1}^{(n)} + \lambda_{k}^{(n)} + \lambda_{k+n+1}^{(n)} + \lambda_{k+n}^{(n)} = c \lambda_{k}^{(n+1)}.
\end{equation*}

%%%%%%% Delta 20 - 22

We solve for the ground states of $H$ first by setting $c=0$. Recalling the dark state condition $\lambda^{(2)}_{k}=0$, the system of ground state equations simplifies as follows: for $k=1,\ldots,N$
\begin{equation*}
    \lambda_{k}^{(l)} + \lambda_{k+1}^{(l)} = 0, \; l=3,\ldots, n
\end{equation*}
and for $k=1,\ldots,n$
\begin{equation*}
    \lambda_{k}^{(n+1)} + \lambda_{k+1}^{(n+1)} = 0.
\end{equation*}
From these simple equations we obtain, for $l=3,\ldots,n$, $\lambda_{k}^{(l)} = (-1)^{k} a_{l}$ for some $a_{l} \in \mathbb{C}$ while periodicity in $n$ of $\lambda_{k}^{(n+1)}$ means that $\lambda_{k}^{(n+1)} = 0$ if $n$ is odd but $\lambda_{k}^{(n+1)} = (-1)^{k} a_{n+1}$ if $n$ is even. We conclude that the two-particle ground states of $H$ which are also dark states are linear combinations of the states $\ket{\Phi_{3}},\ldots, \ket{\Phi_{n}}$ and, if $n$ is even, $\ket{\Psi_{n+1}}$ defined in Eqs.\ \eqref{eq:darkstatestwo1} and \eqref{eq:darkstatestwo2}.
%\begin{equation*}
%    \lambda_{k}^{(n+1)} = \begin{cases} 0 &\mbox{if $n$ is odd,}\\
%    (-1)^{k} a_{n+1} &\mbox{if $n$ is even}\end{cases}
%\end{equation*}
%To conclude, when $N$ is even, there are $2\lfloor N/4 \rfloor - 1$ dark eigenstates of $H$ in $\mathcal{H}_{2}$ corresponding to the eigenvalue $0$.

%%%%%%%% Delta 23 - 

In the remainder of this Appendix we argue why the ground states of $H$ found above are the only dark states in $\mathcal{H}_{2}$. We consider the eigenstates of $H$ in $\mathcal{H}_{2}$ corresponding to nonzero eigenvalues: if $c \neq 0$ the system of eigenvalue equations, written out explicitly, reads
\begin{widetext}
\begin{equation}\label{eq:appendixsyseqn}
\begin{aligned}
    &(l=2)\quad &\lambda_{k}^{(3)} + \lambda_{k-1}^{(3)} &= 0,  &\rdelim\}{11}{6em}[$k=1,\ldots,N$]\\
    &(l=3)\quad &\lambda_{k}^{(4)} + \lambda_{k-1}^{(4)} &= c \lambda_{k}^{(3)}, &\\    
    &(l=4)\quad &\lambda_{k}^{(5)} + \lambda_{k-1}^{(5)} &= c \lambda_{k}^{(4)}, &\\    
    &(l=5)\quad &\lambda_{k+1}^{(4)} + \lambda_{k}^{(4)} + \lambda_{k}^{(6)} + \lambda_{k-1}^{(6)} &= c \lambda_{k}^{(5)},  &\\ 
    & & &\enspace \vdots &\\   
    &(l=m)\quad &\lambda_{k+1}^{(m-1)} + \lambda_{k}^{(m-1)} + \lambda_{k}^{(m+1)} + \lambda_{k-1}^{(m+1)} &= c \lambda_{k}^{(m)},  &\\ 
    & & &\enspace \vdots &\\
    &(l=n)\quad &\lambda_{k+1}^{(n-1)} + \lambda_{k}^{(n-1)} + \lambda_{k}^{(n+1)} + \lambda_{k-1}^{(n+1)} &= c \lambda_{k}^{(n)},  &\\ 
    &(l=n+1)\quad &\lambda_{k+1}^{(n)} + \lambda_{k}^{(n)} + \lambda_{k+n+1}^{(n)} + \lambda_{k+n}^{(n)} &= c \lambda_{k}^{(n+1)}, \quad &k=1,\ldots,n.\\ 
\end{aligned}
\end{equation}
\end{widetext}
This system of $n(N-1)$ equations in the $n(N-3)$ unknowns $\set{\lambda^{(l)}_{k}}$ corresponds to a matrix of coefficients $\mathcal{E}_{N}$ with block structure     
%\begin{equation*}
%\mathcal{E}_{N} = \left(\begin{array}{c|ccccccc|c} 
%        A_{N} &  &  &  &  &  &  &  & \\
%        -c \mathbf{1}_{N} & A_{N} &  &  &  &  &  &  & \\
%        \hline
%         & -c \mathbf{1}_{N} & A_{N} &  &  &  &  &  & \\
%         & A_{N}^{\dagger} & -c \mathbf{1}_{N} & A_{N} &  &  &  &  & \\
%         &  & A_{N}^{\dagger} & &  &  &  &  & \\ 
%         &  &  &  &  \ddots &  &  &  &\\
%         &  &  &  &  &  & -c \mathbf{1}_{N} & A_{N} & \\
%         &  &  &  &  &  & A_{N}^{\dagger} & -c \mathbf{1}_{N} & B_{N}\\
%        \hline
%         &  &  &  &  &  &  & B_{N}^{\dagger} & -c \mathbf{1}_{n}        
%    \end{array}\right),
%\end{equation*} 
\begin{equation*}
\left(\begin{array}{c|cccccc|c} 
        A_{N} &  &  &  &    &  &  & \\
        -c \mathbf{1}_{N} & A_{N} &  &    &  &  &  & \\
        \hline
         & -c \mathbf{1}_{N} & A_{N} &    &  &  &  & \\
         & A_{N}^{\dagger} & -c \mathbf{1}_{N} & A_{N} &  &   &  & \\
         &  & A_{N}^{\dagger} & \ddots &    &  &  & \\ 
%         &  &  &  &     &  &  &\\
         &  &  &    &  & \! \! \! \! -c \mathbf{1}_{N} & A_{N} & \\
         &  &  &    &  & \! \! A_{N}^{\dagger} & -c \mathbf{1}_{N} & B_{N}\\
        \hline
         &  &  &    &  &  & B_{N}^{\dagger} & -c \mathbf{1}_{n}        
    \end{array}\right)
\end{equation*}  
consisting of $n$ block rows and $n-1$ block columns. Here $A_{N}$ is the $N \times N$ circulant matrix with initial $1 \times N$ row $(1, 0, \ldots, 0, 1)$ and $B_{N}$ is the $N \times n$ block matrix $B_{N} = [b_{n}; b_{n}]$ composed of two vertical copies of the $n \times n$ circulant matrix $b_{n}$ with initial $1 \times n$ row $(1,0,\ldots,0,1)$; $\mathbf{1}_{k}$ denotes the $k \times k$ identity matrix. 

The matrix $\mathcal{E}_{N}$ is full rank, and so by the the Kronecker-Capelli Theorem the system of equations \eqref{eq:appendixsyseqn} has no solutions. To see this, we start with the first $N$ columns of $\mathcal{E}_{N}$ and note that the rank of the block $A_{N}$, being a circulant matrix, is $N-1$. The superposition of the first $N$ columns eliminating the columns of $A_{N}$ results in a nonzero column in the (trivially full rank) block $-c \mathbf{1}_{N}$ which cannot be made to vanish using the second set of $N$ columns. This means the first $N$ columns are linearly independent; continuing this process with the second set of $N$ columns, and so on, we find that all columns are linearly independent, and so $\mathcal{E}_{N}$ is full rank. Therefore, there are no nontrivial solutions to the system of eigenvalue equations, and so there are no eigenstates of $H$ in $\mathcal{H}_{2}$ corresponding to nonzero eigenvalues.

For example, consider the case $N=6$: the coefficient matrix $\mathcal{E}_{6}$ reads
\begin{equation*}
\mathcal{E}_{6} = \left(\begin{array}{c|c} 
    A_{6} & 0\\
    -c \mathbf{1}_{6} & B_{6} \\
    \hline
    B_{6}^{\dagger} & -c \mathbf{1}_{3}
\end{array} \right),
\end{equation*}
where
\begin{equation*}
 A_{6} =  \left(\begin{array}{cccccc} 
    1 & 0 & 0 & 0 & 0 & 1\\
    1 & 1 & 0 & 0 & 0 & 0\\
    0 & 1 & 1 & 0 & 0 & 0\\
    0 & 0 & 1 & 1 & 0 & 0\\
    0 & 0 & 0 & 1 & 1 & 0\\
    0 & 0 & 0 & 0 & 1 & 1
\end{array} \right),\quad B_{6} =  \left(\begin{array}{cccccc} 
    1 & 0 & 1\\
    1 & 1 & 0\\
    0 & 1 & 1\\
    1 & 0 & 1\\
    1 & 1 & 0\\
    0 & 1 & 1
\end{array} \right).
\end{equation*}
The rank of $\mathcal{E}_{6}$ is $9$, which is equal to the number of unknown variables, and we conclude that there is no nontrivial solution. 

%Similarly, for $N=10$ the matrix of coefficients reads
%\begin{equation*}
%\mathcal{E}_{10} = \left(\begin{array}{c|cc|c} 
%    A_{10} & 0 & 0 & 0\\
%    -c \mathbf{1}_{10} & A_{10} & 0 & 0\\
%    \hline
%    0 & -c \mathbf{1}_{10} & A_{10} & 0\\
%    0 & A^{\dagger}_{10} & -c \mathbf{1}_{10} & B_{10}\\
%    \hline
%    0 & 0 & B^{\dagger}_{10} & -c \mathbf{1}_{5}
%    \end{array} \right)
%\end{equation*}
%whose rank, $35$, is equal to the number of unknowns meaning there are no nontrivial solutions. 

\end{document}